\documentclass{aa}  

\usepackage{graphicx}
\usepackage{txfonts}
\usepackage{xspace}
\usepackage{gensymb}

\usepackage{natbib}
\bibpunct{(}{)}{;}{a}{}{,} % to follow the A&A style
\newcommand{\req}[0]{$R_{\rm eq}$\xspace}
\newcommand{\porb}[0]{$P_{\mathrm{orb}}$\xspace}
\newcommand{\vp}[0]{$v_{\phi}$\xspace}

\usepackage{ulem}
\usepackage{color}
\usepackage[dvipsnames]{xcolor}

\newcommand{\fix}[1]{\textcolor{black}{{#1}}}

\newcommand{\revi}[1]{\textcolor{black}{{#1}}}

\begin{document}

   %\title{A full description of the structure of coplanar Be binaries}
    %\title{High resolution SPH simulations of Be binaries}
    \title{High-spatial-resolution simulations of Be star disks in binary systems:}

   \subtitle{I. Structure and kinematics of coplanar disks}

   \author{A. C. Rubio\inst{1,2,3},
          A. C. Carciofi\inst{2},
          J. E. Bjorkman\inst{4},
          T. H. de Amorim\inst{2},
          A. T. Okazaki\inst{5},
          M. W. Suffak\inst{6},
          C. E. Jones\inst{6},
          \and
          P. P. Candido\inst{2}
          %\fnmsep\thanks{Just to show the usage
          %of the elements in the author field}
          }

   \institute{Max-Planck-Institut für Astrophysik, Karl-Schwarzschild-Str. 1, 85748 Garching b.\ M\"unchen, Germany \and
   Instituto de Astronomia, Geof{\' i}sica e Ci{\^e}ncias Atmosf{\'e}ricas, Universidade de S{\~ a}o Paulo, Rua do Mat{\~ a}o 1226, Cidade Universit{\' a}ria, 05508-900 S{\~a}o Paulo, SP, Brazil
         \and
    European Organisation for Astronomical Research in the Southern Hemisphere (ESO), Karl-Schwarzschild-Str.\ 2, 85748 Garching b.\ M\"unchen, Germany
         \and
    Ritter Observatory, MS113, Department of Physics and Astronomy, University of Toledo, Toledo, OH43606-3390, USA
        \and
    Center for Development Policy Studies, Hokkai-Gakuen University, Toyohiraku, Sapporo, Hokkaido 062-8605, Japan \and
    Department of Physics and Astronomy, Western University, London, ON N6A 3K7, Canada
    \\
    \email{amanda.rubio@usp.br - rubio@mpa-garching.mpg.de}
             %\thanks{The university of heaven temporarily does not
             %        accept e-mails}
             }

   %\date{Received Dec, 2024; accepted XXX, XXX}

% \abstract{}{}{}{}{} 
% 5 {} token are mandatory
 
  \abstract
   {Over the past decades, binarity in O and B stars has proven to be an important aspect in the their birth, evolution, and death. For the particular case of Be stars, binarity might be the cause for their spin up, and thus part of the puzzle of the mechanism behind their mass outbursts that lead to the formation of their viscous decretion disks. Detecting companions in systems with Be stars can be challenging, making it difficult to obtain observational constraints on the binary fraction of Be stars.}
   {We explore the effects of a binary companion in a system with a Be star, from the moment when the disk first begins to form until it reaches quasi steady-state. The tidal forces considerably affect the Be disk leading to the formation of distinct regions in the system, with observational consequences that can be used to infer the presence of a otherwise undetectable companion. }
   {We use \revi{smoothed particle hydrodynamics} (SPH) simulations of coplanar, circular binary systems \revi{with Be parameters (star mass of 12.9 $M_{\odot}$, equatorial radius of 5.5 $R_{\odot}$, and effective temperature of 26000 K) fixed, and varying the orbital periods (30, 50 and 84 days), disk viscosities ($\alpha = 0.1$, 0.5, and 1) and mass ratios ($q = 0.16$, 0.33, and 0.5).} High \revi{spatial} resolution is achieved by adopting particle splitting in the SPH code, as well as a more realistic description of the secondary star and the disk viscosity.
   }
   {\revi{With the upgraded code, we can probe a region approximately 4 times larger than previously possible. Our models show that the disk can be divided into five regions of interest: the inner Be disk, the spiral dominated disk, the bridge, and previously unseen circumsecondary and circumbinary regions, revealed thanks to the increased resolution of the simulation. We describe the configuration and kinematics of each region, and provide a summary of their expected observational signals. In all simulations there is mass transfer from the Be disk into the Roche lobe of the companion via the bridge. In other words, the disk is not sharply truncated at a given radius, but rather suffers a strong decrease in density in region spanning several stellar radii. This truncation region is  azimuthally variable, elongated along the Roche potentials of the binary system. 
   %\rmACC{Disk material escapes through the bridge to the Roche lobe of the companion}. 
   %\commentACC{Voce já disse o quue está na frase acima. Agora precisamos costurar com o que vem abaixo.}
   \revi{Material that enters the Roche lobe of the companion is partially captured by it, effectively forming a rotationally supported, disk-like structure.}
   Part of the material is accreted by the companion in all simulations, but the expected X-ray emission of this accretion is faint. Material not accreted escapes the Roche lobe of the companion and forms a circumbinary, one-armed spiral around the system. This is the first work to describe the region beyond the truncation region of the Be disk and its observational consequences with detail.} }
   {All five regions are present for all models explored in this work. The effects of orbital period, viscosity, and mass ratio on the structure of Be binary systems are significant, and \revi{should impact} the observables. \revi{Based on our models, we argue that observational features of previously unclear origin, such as the intermittent shell features and emission features of HR\,2142 and HD\,55606, originate in areas beyond the truncation region.} This new understanding of the behavior of \revi{disks in} Be binaries will allow not just for better interpretation of existing data, but also for the planning of future observations.  
   % \commentACC{Nao sei se colocaria esse final, pois voce escolheu apenas uma previsão entre tantas...}
   % Emission and absorption in the bridge, circumsecondary and circumbinary regions are likely responsible for peculiar spectrocopic and interferometric signals seen in Be+sdO binaries HR\,2142 and HD\,55606. Such observations have therefore diagnostic power to derive properties not only of the companion and orbit, but also of the viscosity of the Be disk.
   }


    \titlerunning{High-spatial-resolution simulations of Be star disks in binary systems}
    \authorrunning{A. C. Rubio, A. C. Carciofi, J. E. Bjorkman, et al. }

   \maketitle
%
%________________________________________________________________

% \textcolor{red}{
% Pontos a abordar:
% \\
% - Na discussão da taxa de acresção na secundária, comparar os valores obtidos com valores típicos de Binárias de raios X
% }

% \ACR{Falar de HR2142}

\section{Introduction}\label{sec:intro}

% \commentACC{Resolver antes de submeter: 1) ortografia está misturada. Tem que ser inglesa ou americana, de forma consistente. Sugiro americana pois estamos mais acostumados. 2) Sempre que possível substitua as sentenças na voz passiva por voz ativa. Isso é fortemente recomendado pelo guia do autor.}

From protostars to black holes, disk structures are common in astrophysical systems. In the disks of mass transfer binary systems, young stellar objects, and active galactic nuclei, viscosity is the main property that leads to the fluid forming a disk in response to various forces, including gravity, gas pressure, and radiative pressure.
These types of viscosity-driven disks are commonly referred to as \textit{$\alpha$-disks}, from the formulation of \citet{shakura1973}. Most $\alpha$-disks are accretion disks, built from outside-in (such as the disks of young stars, mass transfer binaries, and AGN).
An exception to this are the decretion disks of classical Be stars, that are built inside-out from material ejected from the central star. A hypothesis is that a combination of non-radial pulsations (NRPs) and the ubiquitous fast rotation of Be stars is responsible for the mass ejection events that form the disk \citep[see][for an overview of this and other proposed scenarios for the formation of Be disks]{owoki2006}. 
%\commentACC{concordo com o Tajan. O artigo do Owocki é uma boa}
These disks, whose presence lead to linear polarization, \revi{infrared (IR)} excess, and Balmer lines in emission \citep{rivinius2013}, are intermittent: they are maintained while mass is being ejected from the central star, which happens episodically and unpredictably for most Be stars, but dissipate when the mass ejection ceases. Their build-up and dissipation phases happen in human timescales (months--years), making Be stars engaging targets to explore disk physics \citep[e.g.,][an overview of the very active Be star $\gamma$ Cas]{baade2023}.

%\ACRtd{(CITE THE GCAS PAPER?)}. 
%\commentACC{Em que contexto? Não vi contexto aqui para citar. Poderia terminar citando o review aqui, que é meio que um coringa nestes casos.}\answerACR{no contexto de gCas passar por várias fases de disco muito complexas}
%\commentACC{OK}

Rapid rotation is necessary for a B star to become a Be star. \revi{Some studies indicate that the threshold decreases with spectral type, being as low as 0.4 of critical rotation for early-type (more massive) Be stars (\citealt{cranmer2005}, \citealt{huang2010}). Such correlation is, however, tentative, as both the rotational velocity and the spectral type (based on effective temperature and log(g) estimates) are difficult to measure in Be stars \citep{meilland2012}}. How Be stars acquire such high rotational speeds is still debated in the literature. \revi{There are three proposed pathways: i) they are simply born as fast rotators, and continue to be so throughout the MS \citep{bodenheimer1971}; ii) they increase their surface rotation rate during the MS evolution as the stellar core contracts and angular momentum is redistributed within the star \citep{ekstrom2008, georgy2013}; and iii) they are formed in MS+MS binaries, gaining angular momentum through mass transfer \citep{pols1991}. It is likely that all three play significant roles in creating the population of fast rotating B stars that can become Be stars. Nevertheless, the} notion that Be stars are survivors of mass transfer binary systems has strengthened considerably in recent years \citep{demink2013, boubert2018, hastings2021, dallas2022, dodd2024}. 
If this scenario is true for at least a portion of Be stars, then they should be found in systems with evolved objects such as white dwarfs (WD), neutron stars (NS), black holes (BH), and stripped stars such as O and B subdwarfs (sdO/sdB) \citep{rivi2024}.
%\commentACC{CITAR ARTIGO DO RIVI QUE FOI RECENTEMENTE PUBLICADO (ou aceito)}
%\commentACC

%Be stars are not rare: the fraction of Be stars among B stars is 20-30$\%$ \citep{zorec1997} (CHECK THIS. IN THE GALAXY??). And, 
It is not rare to find Be stars with companions. As massive stars, they are common in binary systems \citep{sana2012}, and most high mass X-ray binaries ($\sim$60$\%$ - \citealt{liu2006}) fall into the subclass of the Be/X-Ray binaries (BeXRB), where the primary is a Be star. X-ray emission occurs as matter from the decretion disk is accreted by the secondary. There are 81 confirmed BeXRBs in the Galaxy, 48 of which are \revi{known} to host a NS companion \citep{shao2014, reig2011, liu2006}.

% When the companion is a star that has been stripped of its envelope (e.g., O and B subdwarves -- sdO/sdB), their high temperatures {lead to enhanced emission in the far ultraviolet (FUV)}.

{After mass transfer, the remaining exposed core of the donor is generally not observable in the visible, \revi{which is} dominated by the brighter Be star. In the case of subdwarfs, however, their high temperatures lead to enhanced emission in the far ultraviolet (FUV).}
% \commentACC{concordo com Tajan}
%\commentACC{Não conheço nenhum caso em que a estrela fica mais brilhante no FUV. Lembre-se que corpos negros não se cruzam. O mais correto é dizer que o brilho relativo aumenta no FUV}
Spectral disentangling of FUV spectra of Be stars has allowed for the detection of such systems, for example $\phi$ Per \citep{gies1998}, FY CMa  \citep{peters2008}, 59 Cyg \citep{peters2013}, HR2142 \citep{peters2016}, 60 Cyg \citep{wang2017}, and the candidates found by \citet{wang2018, wang2021}. Interferometry has also become an effective tool for detecting these companions and characterizing their orbits. The recent work of \citet{klement2022} was the first to detect an sdB companion, in the $\kappa$ Dra system, while the lack of detection of sdO/B companions in $\gamma$ Cas analogues\footnote{$\gamma$ Cas, FR\,CMa, HR\,2370, V558\,Lyr, V782\,Cas, and $\pi$ Aqr} by \citet{klement2024} points to them being Be+WD systems.

% The companion of a Be star will impact the Be decretion disk should the orbital separation be sufficiently small. 
\revi{The companion of a Be star can affect the Be decretion disk if the orbital separation is small enough to allow significant gravitational and tidal interactions.}
Hot companions can irradiate the disk, increasing its temperature locally and altering the strength of recombination lines formed in the disk, such as Balmer lines (\citealp[HR\,2142 --][]{peters2016}, \citealp[HD\,55606 --][]{chojnowski2018}). 
Moreover, tidal interactions with the companion have the potential to markedly alter the form and dynamics of the Be disk \citep[e.g.,][]{chojnowski2018}. The works of \citet{okazaki2002} and  \citet{panoglou2016} used a 3D smoothed particle hydrodynamics code (SPH -- \citealt{benz1990}) to study Be binary systems in coplanar orbits.
\citet{cyr2017} and \citet{suffak2022} did the same for misaligned systems. Their results show that the most significant consequences of the secondary/disk interaction are the emergence of two-armed density waves \revi{\citep[composed of two modes, thus $m=2$,][]{okazaki1991}} lead by the secondary, the accumulation of matter in the inner disk caused by the tidal torque {hindering the outward transport of angular momentum}, and the truncation of the disk. The disruption of the disk is even more complex for misaligned systems: the tilted disk can be warped, tearing in more extreme cases, and subject to Kozai-Lidov oscillations
\citep{suffak2022}.

%The SPH code uses an ensemble of low mass particles to simulate a fluid and thus solve the hydro-dynamical equations for this fluid in a time-dependant manner. 

%\commentACC{Voce falou do m=2 e truncation, tendo contrapartidas observacionais. Por completeza, precisa falar do disk tearing, que foi sugerido por Marr et al. para explicar o comportamento de Pleione, e recentemente ganhou mais suporte observacional do Marr (ultimo artigo dele)}

The two-armed density waves have clear observational counterparts for several binary Be systems. In general, Be stars have double-peaked Balmer lines in emission due to the Doppler effect of disk rotation (when the system's orientation is not exactly pole-on). As the waves travel through the disk, the density is temporarily increased in one of its sides alternately. Thus, the relative strengths of the red (R) and violet (V) sides of the emission lines formed in the disk (specially H$\alpha$) vary cyclically. A very clear example of this is seen in the binary Be star $\pi$ Aqr, as shown by \citet{zharikov2013}, where the violet to red (V/R) variations in H$\alpha$ have the same period as the orbit of the companion ($\sim84$ days).

Disk tearing, the more extreme of the consequences of binary interaction with Be disks in misaligned systems, will also have observable consequences. The recent work of \citet{marr2022} explains the complex spectroscopic and polarimetric behavior of the Be star Pleione (HD\,23862, 28 Tau) in the context of disk tearing. They suggest that Pleione's disk is first tilted then torn into an outer and an inner disk by the torque exerted by its companion. The inner disk returns to the orbital plane, while the outer disk precesses and finally dissipates after a period of 15 years. The whole disk tearing process is repeated every 34  years, explaining the shifts to shell emission lines and polarization angle seen in the data.
This scenario has gained further qualitative support by the recent work of \citet{2024MNRAS.527.7515S} who presents \textsc{hdust} \citep{carciofi2006} radiative transfer calculations based on the models of \citet{suffak2022}, demonstrating that disk tearing indeed produces variability consistent to what is observed in Pleione.

Disk truncation is also directly correlated to an observational effect. The concept has been used by \citet{klement2017, klement2019} to explain a prevalence of a steepening in the slope of the spectral energy distribution (SED) towards radio wavelengths \citep[dubbed SED turndown by][who first noticed this behavior]{waters1991} in their \revi{sample of Be stars. A truncated disk, having a smaller emitting area, has a lower than expected emission in larger wavelengths when compared to disks of isolated Be stars \citep{vieira2015a}.} \citet{klement2019} notes, however, that the radio SED slopes for some of their sample are inconsistent with a true disk truncation, i. e., the scenario where effectively no significant trace of a disk exists beyond the orbit of the secondary does not befit the data. This indicates that a circumbinary disk is present in these systems, and that its density is non-negligible. This circumbinary structure was completely absent in the SPH simulations calculated by \citet{okazaki2002}, \citet{panoglou2016}, and all other works using this same SPH code, \revi{but was formed in the simulations of \citet{martin2024}}.

In this work, we present new SPH simulations calculated with an upgraded version of the SPH code of \citet{okazaki2002}. For the first time we are able to follow the entire evolution of the disk of a Be star in a coplanar, circular binary. \revi{Our goal is to offer a full description of the behavior of Be disks in these conditions, thereby bridging theory and observations. 
In Sect.~\ref{sec:diskphysics} we present an overview of current paradigm for Be disks, the improved SPH code and the modifications we implemented. In Sect.~\ref{sec:models} we describe our simulations, presenting a complete overview of the full structure of the perturbed disk, including the area around and beyond the secondary. Each region has its own dedicated subsection, where we also discuss their observational consequences, and how they can be used to infer the presence of an otherwise hidden companion. An important note is that our simulations are purely hydrodynamical and isothermal. Including radiative transfer is the next step, but was beyond the scope of the present work. Thus, the observational expectations described here are based on the dynamics and density structure of the perturbed disk only. The temporal evolution of the disk is discussed in Sect.~\ref{sec:temporal_evol}, and our conclusions are in Sect.~\ref{sec:conclusions}. }

\section{Modeling tools}\label{sec:diskphysics}

% \commentACC{Suggestion for new sect. title and new subsection}

% \commentACC{No geral esta seção está boa, mas o texto ainda precisa ser polido, para evitar repetições e estar estruturado de forma mais linear}

\subsection{Short overview of the VDD model}
\label{sec:vdd_overview}

The current most favored model for the disks of Be stars is the Viscous Decretion Disk (VDD) model (\citealt{lee1991}, further developed by \citealt{bjorkman1997}, \citealt{okazaki2001}, and \citealt{bjorkman2005}), that has been tested extensively in the past decades. The VDD model is a modification of the standard $\alpha$-disk formulation of \citet{shakura1973}, where viscosity, $\nu$, is parameterized in terms of a dimensionless parameter $ 0 < \alpha \lesssim 1$, where $\nu = \alpha c_s H$ \citep[][where $H$ is the scale height of the disk and $c_s$ is the sound speed in the disk]{shakura1973}. The higher the value of $\alpha$ the higher the viscosity of the disk, and therefore the faster the disk evolution. The $\alpha$-disk formulation is commonly applied, with some variations, to the accretion disks in black hole accretion (the context on which it was originally developed), mass transfer binaries, disks of young stellar objects, protoplanetary disks, and even active galactic nuclei disks \citep[e. g.][]{king2007, zhu2009, armitage2011, speri2023}; the disks of Be stars, however, are \textit{de}cretion disks: the source of the mass and angular momentum injected into the disk is the central star \citep[e.g.,][]{rimulo2018}.
% Be stars are all rapid rotators and recent studies suggest they are also all non-radial pulsators \citep[][and references therein]{labardie-bartz2022}. The combination of rotation and pulsation is likely the cause of the mechanism that drives the mass loss of Be stars that leads to the formation of the decretion disk \citep{rivinius1998a}. 
% \commentACC{vdd... acho que podemos simplesmente tirar}
The mass loss of Be stars is usually episodic and in most cases unpredictable. A Be star can spend decades with a steady mass ejection rate feeding its disk
\citep[$\beta$\,CMi being a prime example, e.g.,][]{klement2015}
only for it to suddenly stop; as there is no longer a source of mass and angular momentum being fed to it, the disk dissipates in time scales ranging from weeks to years \citep{rimulo2018}. %Thus, a Be star can be said to be ``active'' when it either has or is building up a disk, or ``inactive'', when no sign of disk or significant mass ejection events are observed.  
%\citet{rimulo2018} and \citet{ghoreyshi2018} use \textsc{SINGLEBE} and the 3D radiative transfer code \textsc{HDUST} \citep{carciofi2006b, carciofi2008b} to further explore these disk build-up and dissipation events, modeling lightcurves of Be stars. They obtained estimates of the viscosity parameter $\alpha$ for build up and dissipation phases for their sample, consistently finding that $\alpha$ varies between the phases, and between disk events even for the same Be star. 

% \commentACC{Eu inverteria a ordem abaixo. Comece discutindo o que o equilibrio hidro. implica, chegando na formula. Essa formula preve algulos de poucos graus perto da estrela e maiores fora, em concordancia com as observacoes.}

After mass is ejected, it is the viscous torque $\nu$ that allows the disk to grow.
%\commentACC{define H, cs}
If mass is ejected from the Be star at a constant rate and symmetrically from its equator, the resulting disk will have non-zero net motion in the azimuthal and radial directions, reaching a quasi-Keplerian velocity field \citep{krticka2011}, but can be considered to be in hydrostatic equilibrium in the vertical direction. 
Following \citet{bjorkman2005}, the surface density structure of the viscous disk can be written as

\begin{equation}\label{eq:VDDsigmabig}
    \Sigma(r) = \frac{\dot{M}_{\mathrm{inj}} v_{\rm orb} R_{\star}^{1/2}}{3 \pi \alpha \, c_s^2 \,r^{3/2}} \left[\left(\frac{R_0}{r}\right)^{1/2} - 1 \right]\,,
\end{equation}

\noindent where $\dot{M}_{\mathrm{inj}}$ is the mass injection rate into the disk, $R_{\star}$ is the stellar equatorial radius, $v_{\rm orb}$ is the Keplerian orbital velocity at the equator, $R_0$ is an integration constant, and the sound speed in the disk is $c_s = ({kT/m_\mathrm{H} \mu})^{1/2}$ ($T$ is the disk temperature, $k$ is the Boltzmann constant, $m_\mathrm{H}$ is the hydrogen mass, and $\mu$ is the mean molecular weight of the gas). 

%\commentACC{citar kurfurtz}

Since the VDD is in hydrostatic equilibrium, if we assume isothermality, its scale height $H$ is controlled only by the gas pressure and the gravitational force of the star. Therefore
\begin{equation}\label{eq:scaleh}
    H = H_0 \left( \frac{r}{R_{\star}} \right)^{3/2},
\end{equation}
where $H_0 = c_s R_{\star}/v_{\mathrm{orb}}$. Thus, the disk flares as it grows and distances itself from the central star, as observations indicate (\citealt{quirrenbach1997}, \citealt{wood1997}, \citealt{hanuschik1996}, \citealt{carciofi2008}).
{Equation~\ref{eq:VDDsigmabig} holds under the thin disc approximation, which is only valid near the star. As the disk flares and $H << r$ is no longer applicable, the thin disk approximation becomes invalid \citep{kurfurst2018}.}

% \commentACC{Amanda, o acima está estranho. No final do 2o paragrafo acima, a gente fala do flare to disco no contexto da aproximacao de disco fino. Mas isso só é definido no paragrafo seguinte. Acho quem uma reorganização é necessária }

With these assumptions we can express the surface and volume density of a steady-state VDD simply as
\begin{equation}\label{eq:VDDsigma}
    \Sigma(r) = \Sigma_0 \left( \frac{r}{R_{\star}} \right)^{-m} ,
\end{equation}

\begin{equation}\label{eq:VDDrho}
    \rho(r, z) = \rho_0 \left( \frac{R_{\star}}{r} \right)^{-n} \exp  \left( \frac{-z^2}{2H^2} \right), 
\end{equation}
where the radial density exponents $m$ and $n$ are equal to 2 and 3.5, respectively, as \revi{for an isothermal disk, $m$ and $n$ relate to each other as $n = m + 1.5$ \citep{bjorkman2005}.}
However, the above is often only a rough representation of real Be disks.

Most Be stars undergo complex phases of disk build-up and dissipation, resulting in inherently complex disks. These mass loss rate variations can be gleaned from long-term photometric monitoring of Be stars \citep[e.g.,][]{2002AJ....124.2039K}.
Much effort has been put in modeling these more active Be stars. \citet{haubois2012, haubois2014}, for instance, studied the dynamical evolution the disk in various feeding scenarios, using the 1D, time-dependent hydrodynamic code \textsc{SINGLEBE} \citep{okazaki2007}. 
% \commentACC{Cuidado aqui, reescrever. o Haubois analisou a fotometria do ponto de vista teórico, principalmente. A comparação com dados é só uma pequena parte do artigo, e ainda asism tudo qualitativo}
% \commentACC{reescrever o abaixo, está meio circular}
They find that the $\rho \propto r^{-3.5}$ solution is only achieved when the disk feeding rate $\dot{M}_{\mathrm{inj}}$ is kept (nearly) constant for a sufficiently long time. In their models, values of $n < 3.5$ are only reached when the disk is in dissipation, while $n > 3.5$ are attributed to disks in build-up stage. However, in their mid-infrared survey of 80 Be stars, \citet{vieira2017} found several Be stars with $n < 3.5$, even though some of these stars have had a spectroscopically stable disk for decades. 
This contradiction to theory demands an explanation and appears to result from a combination of factors whose respective significance has yet to be thoroughly investigated.
One relevant factor, frequently left aside in models, is that the gas is non-isothermal. \citet{carciofi2008}, for instance, found that, near the star, the electron temperature at the midplane falls quickly with distance, causing the density slope to be substantially lower than the predicted $n = 3.5$. \revi{In most cases, the density slope of a real Be disk will not be well described by a single value of $n$ due to a combination of the aforementioned factors. This simplification is nonetheless useful to characterize these disks and their evolution, and widely used in the Be star community.}

% The importance of $\alpha$ and of the mass and angular momentum injection rates in the dynamics of Be disks is made clear by their results. que frase 'obvia hein kkkkkk
% \commentACC{Aqui tem mais coisas para serem ditas e que sao importantes no contexto do artigo. Por exemplo, nos modelos dele n < 3.5 so ocorre com discos em dissipação, o que levou Vieira 2017 a classificar como tal as estrelas da sua amostra com outros valores. }

{Another effect that can impact $n$ is related to the presence of a binary companion, and was only discovered when SPH models of Be binary systems became available.}
An SPH code \citep{lucy1977} uses an ensemble of thousands of particles to simulate the behavior of a fluid. It solves the hydrodynamic motion equations for each particle as discrete points, and then sums over the ensemble using a kernel function interpolator to obtain smoothed approximations of the physical quantities of the fluid. \citet{okazaki2002} adapted the SPH code of \citet{benz1990} and \citet{bate1995} to simulate single and binary Be stars. Their initial results were of great relevance: the simulations confirmed that the presence of a companion leads to disk truncation (as previously proposed by \citealt{okazaki2001a}), and find that the disks of binary Be stars are denser than those of isolated Be stars, as the torque exerted by the companion prevents matter and angular momentum from spreading outwards, an effect that is more pronounced for lower viscosity. This effect was later dubbed `accumulation effect' by \citet{panoglou2016} and causes the density slope to be smaller than the predicted $n = 3.5$ value for isothermal disks.

\citet{panoglou2016} studied the disk dynamics of coplanar Be binaries using the same SPH code \revi{while harnessing advancements in computational power since 2002}. Their results further confirm the disk truncation and the accumulation effect caused by the tidal interaction of the companion with the Be disk. They find also that two-armed density waves are excited in the disk by this tidal interaction, which also deforms the disk: while they are axissymmetric for the isolated case, they become elongated under the influence of the companion. \citet{panoglou2016} ran several identical simulations, but with different values of $\alpha$, orbital period ($P_{\mathrm{orb}}$) and mass ratio ($q$), so that their individual effects could be probed. They find that the tidal effects of the secondary are enhanced when $q$ is larger, or \porb smaller, as both increase the gravitational and tidal influence of the secondary on the Be disk: the disk becomes smaller, denser, and more oblate. Higher values of the viscosity parameter and of \porb lead to less deformed, bigger disks. {The subsequent works of \citet{cyr2017} and \citet{suffak2022} explore the effects of different orbital configurations in Be binaries, finding that the companion can also tilt and even tear the Be disk.}

All the previously mentioned works use the same SPH code \citep{okazaki2002}. %, in which the Be decretion disk is built from first principles, growing throughout the simulation. The Be star and its companion are ``sink particles'', which can accrete the gas particles that compose the disk. 
%As described in Sect.~\ref{sec:diskphysics}, the density profile of a Be disk falls off with radius, i.e, the inner disk is much denser than the outer disk. 
In a SPH code, the resolution of a region depends on the number of particles in that region. As all gas particles have the same mass, less dense regions tend to have a lower resolution. Therefore, the models in previous works achieved good resolution in the inner disk, but had few particles in the outer regions, as the density profile of a Be disk falls off with radius (see Sect.~\ref{sec:diskphysics}). The resolution of the models in these works was adequate for their propose, which was to describe the effects of the companion on the structure and kinematics of the Be disk. To further save computing power, these works also simplify the region around the companion by making the accreting particle representing the companion as large as its \revi{Roche lobe (RL)}. All particles that entered that region were, therefore, immediately removed from the simulation. This, combined with the inherently low resolution of the outer disk, results in a Be binary where only the region within the Roche lobe of the Be star is resolved, providing no insight into the conditions around and beyond the companion. To address these issues, we modified the code of \citet{okazaki2002}, as described in the next section.

% A way to increase the number of particles in low density regions is to decrease the particle mass while maintaining the mass injection rate. However, for a disk with a density structure like Be disks, this solution is computationally very costly, as it would increase the number of particles in the entire simulation, leading to an extremely over-resolved inner disk.

% \commentACC{Amanda, ao chegar aqui sinto que faltou coisa, pois abaixo voce já está entrando nos detalhes do código e o leitor vai ficar se perguntando para onde isto vai. Acho que aqui voce deve continuar revisando em detalhes os principais resultados SPH e concluir falando dos seus problemas que voce resolve aqui. Fazer de forma curta, pois será retomado abaixo }

\subsection{Improved SPH code }\label{sec:sphcode}

% \commentACC{revisar, há alguma repetição com o texto acima}

Three critical modifications, designed to resolve the issues highlighted in the preceding section, were applied to the SPH code of \citet{okazaki2002}. Below we provide only the code details relevant for this work. For more comprehensive information, interested readers are directed to \cite{okazaki2002} and references therein.

% \commentACC{Faltou: definir o que é smoothing length e seu símbolo h. Voce usa o simbolo abaixo sem defini-lo. Acho que é bom também colocar a formula que o código usa.} \answerACR{A fórmula do kernel? não vejo a necessidade.}

 %For more details we refer to \citet{okazaki2002}.

% \commentACC{Sugiro remover as referencias, mantendo uma descricao genérica}
The code simulates Be binary systems by creating a central point mass sink particle to represent the Be star and then artificially inserting a given number of gas particles in a low orbit around the equator of the sink particle at every 4 time-steps of the simulation. The injection zone is a ring around the stellar equator at 1.04 $R_{\star}$.
This creates a constant mass ejection that feeds the disk for the whole simulation (the dissipation of the disk will be studied in a future publication). %As such, the Be disk is created from scratch and builds-up naturally.
Another point mass sink particle is used to represent the binary companion. \revi{Sink particles are massive points in the simulation act as a gravitational well.} Gas particles that fall into these sink particles are accreted, removed from the simulation. %\revi{The sink particles of the Be star and the companion are the only relevant sources of gravitational force in the simulation.} 
The main parameters of the code are the mass and equatorial radius of the two stars ($M_1$, $M_2$, \req and $R_2$), the mass injection rate, $\dot{M}_{\rm{inj}}$, the orbital period of the system, $P$, the Be star effective temperature, $T_{\rm{eff}}$, and the viscosity parameter, $\alpha$. The code uses a variable smoothing length ($h$) to calculate smoothed quantities around each region, so that the number of neighbor particles is kept {approximately} at 40. \revi{Lower number of particles lead to higher statistical errors in the calculations of smoothed quantities \citep{bate1995}.}
%\commentACC{Sempre? Maximo?}\answerACR{aproximadamente em 40, mas o h tem um tamanho máximo e mínimo tmb}
It has a tree structure to calculate the number of neighbors, and a cubic-spline kernel is integrated with a second-order Runge–Kutta–Fehlberg integrator. The integration is done for each particle at each timestep. The code includes SPH artificial viscosity, variable in space and time to keep the \citeauthor{shakura1973} $\alpha$ fixed for all our simulations. %\commentACC{reference for the SPH artificial viscoisyt} \answerACR{tá td no paper do Okazaki mesmo}

\subsubsection{Secondary's sink}\label{sec:sink}

Previous works with this SPH code were focused on the Be disk, not on the companion. 
%The computational cost of achieving a high-resolution depiction of the interface between the disk and the secondary star, and the circumbinary region, was too high. 
These authors chose to designate the secondary's sink radius to match its \revi{RL} size rather than its stellar radius. The larger sink radius means that all particles that fell into the RL of the companion, a region that would be unresolved in either case due to the small number of particles in the region, %and whose $h$ would be larger than the size of the star itself,
were accreted. Thus, the connection between the companion and the Be disk, and inside of the RL of the companion were not explored by these simulations.
% \commentACC{Este paragrafo repete muito algo que foi dito acima...}
% \commentACC{Sobre o comentário do Tajan, concordo que é bom tentar explicar melhor isso}
% \answerACR{dei uma reescrevida aqui}

%he works of \citet{okazaki2002} and others had characterising the Be disk and exploring the effects of the secondary on it as main goals. 
%Hence, there was no need to achieve a high-resolution depiction of the interface between the disk and the secondary star, or the circumbinary region. 
%\commentACC{VER PONTO AAA ABAIXO}
% These regions, as previously mentioned, have inherently lower density, and consequently, lower resolution in their SPH simulations. 
% This was aggravated by the fact that 
%these authors designated the secondary's sink radius to match its Roche lobe (RL) size rather than the stellar radius. 

Following \citet{eggleton1983}, the RL size is given by

\begin{equation}
    R_L = \frac{0.49 q^{2/3}}{0.6 q^{2/3} + \ln(1 + q^{1/3})}\,,
\end{equation}

\noindent where $q$ is the mass fraction $M_2/M_1$ \revi{and the separation between the stars is normalized to 1}. This is the most used RL approximation, but when $0.1 \lesssim q \lesssim 0.8$, one can further approximate it to simply \citep{accretionpowerinastrophysics}
\begin{equation}
    R_L = 0.462 \left( \frac{q}{q + 1} \right)^{1/3} .
    \label{rochelobe}
\end{equation}
Consequently, the size of sink of the secondary is given by
\begin{equation}\label{eq:rlsize}
    S_2 = f \times R_L = f \times 0.462 \left( \frac{q}{q + 1} \right)^{1/3} ,
\end{equation}
where $0 < f \leq 1$. The sink sphere is centered on the position of the secondary. Previous works, which used this same SPH code, always used $f=1$. 
%\commentACC{Expandir aqui. Primeiro conclua as consequencias das escolhas acima: Having a large sink radius causes many more particles to be absorbed etc. }
This parameter is therefore variable in our simulations so that the sink size matches exactly the size of the secondary.
This change has substantial consequences for the simulations, as explored throughout this paper.

\subsubsection{Particle splitting}\label{sec:psplit}

% \commentACC{AAA: Acho que como trocou coisas de lugar, o texto não está 100\% consistente (tem um as previously mentioned acima que se refere na verdade ao texto abaixo). Eu acho que parte do texto abaixo que explica a baixa densidade deve estar acima, pois compoe o problema descrito na secao anterior. Ao fazer esta mudanca, altere o começo desta seção para não haver repetição.}

In a SPH code, the resolution of a certain region depends on the number of particles present. As all particles have typically the same mass, less dense regions have a lower resolution. Decreasing the particle mass while maintaining the mass injection rate will increase the number of particles in the simulation, and thus its overall resolution. However, for a disk with a density structure like Be disks, where %$\rho \propto r^{-3.5}$, 
$\rho$ usually decreases fast away from the star,  decreasing the mass of each particle is unreasonable: while it would help increasing the resolution in the outer parts, % increase the resolution in the outer disk, 
the number of particles in the inner disk would also increase, 
%The majority of the particles in the simulation are already naturally located in that region. Increasing their numbers will
leading to an over-resolved inner disk and a waste of computational power.

An option that has been used in other SPH works for systems with a large density span is particle splitting. This method allows for particles to split into ``child'' particles that have a fraction of the parent particle mass. The splitting only happens when a particle falls into certain criteria, thus increasing resolution in regions determined by the user. Our implementation follows the work of \citet{kitsionas2002}.

The adopted criteria for splitting a particle are:
\begin{enumerate}
    \item If the particle is inside the Roche lobe of the companion, it splits.
    \item It may still split if it is outside the RL, if all following are true:
    \begin{enumerate}
    \item the particle must be at least $10\,R_{\star}$ away from the Be star, to avoid splits in the already well resolved inner disk;
    \item the particle must have less than 70 neighbor particles, 
    %\commentACC{e o 40 lá de cima?},
    so that the split only happens in low resolution regions; 
    \item the mass of the parent particle cannot be smaller than $4 \times 10^{-4}$ times the initial particle mass (assigned to all particles at the start of the simulation). This avoids that particles in low resolution areas \textit{over}split.
    % \commentACC{Amanda, na pratica isso permite que particulas sejam dividias n vezes, mencionar isso abaixo quando a massa da particula filha for definida}
    \item the smoothing length squared must be larger than 10$\%$ the distance of the particle to the center of mass. This ensures that particles with an $h$ comparable to the disk scale height split.
    \end{enumerate}
\end{enumerate}

% \commentACC{Falar que os critérios acima foram escolhidos de forma a maximizar o número de particulas onde havia baixa resolução, mantendo o número total de particulas -- e o tempo de execução -- o menor possível}

The criteria were chosen so that splitting only happens in the outer disk, most often close to the secondary, increasing the resolution in the less dense regions of the system, but keeping the running time of the simulation reasonable. If a particle fits all requisites, it splits into 13 child particles. The children each have 1/13\textsuperscript{th} of the mass of the parent, with their smoothing lengths being $h \times n^{-1/3}$, where $1 < n < 13$. {Given the constraint on the minimal particle mass set in the conditions above, a particle can only split 4 times consecutively}. Their new positions are at each of the 12 vertices plus the center of an icosahedron with sides 1.5 times the smoothing length of the parent particle, centered on the parent's original position. The velocities and all other properties are retained, copied from the parent to the children.
% \commentACC{Voce achou essa receita em algum artigo? Ou ao menos uma parecida? Se sim, colocar aqui a fonte} \answerACR{Sim, está no text acima, antes da lista com as condições: \citet{kitsionas2002}}

\subsubsection{Viscosity prescription}\label{sec:viscpresc}

Artificial SPH viscosity is the source of the viscous force in the code that allows for the Be disk to grow, following\\
\begin{equation}
  \Pi_{ij}=
    \begin{cases}
      (- \alpha_{\rm SPH} c_s \mu_{ij} + \beta_{\rm SPH} \mu_{ij}^2/\rho_{ij}), & \mbox{if $v_{ij} \cdot r_{ij} \leq 0$}.\\
      0, & \mbox{otherwise} ,
    \end{cases}
\end{equation}
\noindent
where $i$ and $j$ denote two particles, $\alpha_{\rm SPH}$ and $\beta_{\rm SPH}$ are the linear and non-linear artificial viscosity parameters, respectively. The density is $\rho_{ij} = (\rho_i + \rho_j)/2$, the velocity is $v_{ij} = v_i - v_j$, and $\mu_{ij} = h_{ij} v_{ij} \cdot r_{ij} /(r^2_{ij} + (0.01 h^2)^2)$, where $r_{ij}$ is the radial distance between the two particles, and the mean smoothing length is $h_{ij} = (h_{i} + h_{j})/2$ \citep{monaghan1983}. 
% \commentACC{Definir quantidades que faltam ser definidas. REFERENCIA PARA O ACIMA}

This artificial viscosity relates to the more widely used 
%Shakura-Sunyaev 
\citet{shakura1973}
viscosity, $\alpha$, as 

\begin{equation}\label{eq:alpha}
    \alpha = \frac{1}{10} \alpha_{\rm SPH} \frac{h}{H},
\end{equation}
\noindent
where $H$ is the local scale height. In other words, to keep $\alpha$ constant in time and space, $\alpha_{\rm SPH}$ is variable in the simulation according to the scale height and smoothing length, and $\beta_{\rm SPH} = 0$. As our simulations are of binaries, both stellar components must be considered in the calculation of the scale height, otherwise $\alpha$ will nonphysically increase close to the companion. This difference in viscosity becomes relevant now that we have adjusted the size of the sink of the secondary to explore the circumsecondary region. Thus, we have modified the implementation of $\alpha$ as follows.

The vertical force due to gravity acting on the gas, considering the two stars, is
\begin{equation}
    f_z = - \frac{G M_1 z}{(r^2 + z^2)^{3/2}} - \frac{G M_2 z}{(s^2 + z^2)^{3/2}} ,
\end{equation}
where $s$ is the distance from the particle to the secondary projected on the orbital plane, and $M_1$ and $M_2$ are the masses of each star. Following \citet{bjorkman2005}, the equation describing the hydrostatic equilibrium in the vertical direction can be written as (see their Eq.~1 to 11) 
\begin{equation}\label{eq:partial}
    \frac{1}{\rho}\frac{\partial P}{\partial z} = f_z\,,
\end{equation}
assuming axisymmetry and circular orbits for the particles. For an isothermal fluid, where $P = \rho c_s^2$, Eq.\ref{eq:partial} is integrated vertically to give
\begin{equation}\label{eq:integ}
    \ln \left(\frac{\rho}{\rho_0}\right) = \frac{G}{2 c_s^2} \left[ \frac{M_1}{r} \left( \frac{1}{\sqrt{1 + z^2/r^2}} - 1 \right) + \frac{M_2}{s} \left( \frac{1}{\sqrt{1 + z^2/s^2}} - 1 \right) \right] ,
\end{equation}

% \begin{equation}
%     ln (\frac{\rho}{\rho_0}) = \frac{1}{(2 c_s^2)} \left[ \frac{GM1}/{r}  \frac{1}{\sqrt{1+\frac{z^2}{r^2}} - 1) + \frac{GM2}{s} \frac{1}{\sqrt{1+z^2/s^2}} - 1)\right]. 
% \end{equation}
\noindent where $\rho_0$ is the volume density in the equatorial plane. With $r, s \gg z$, Eq.~\ref{eq:integ} becomes
\begin{equation}\label{eq:ln}
     \ln \left(\frac{\rho}{\rho_0}\right) = - \frac{z^2 k}{2 c_s^2}\,,
\end{equation}
where
\begin{equation}
k = G \left( \frac{M_1}{r^3} + \frac{M_2}{s^3}\right)\,.
\end{equation}

%The volume density $\rho$ relates to $P/\rho = c_s^2$, so
Rewriting Eq.~\ref{eq:ln} gives
\begin{equation}
    \rho = \rho_0 \exp{\frac{-z^2 k}{2 c_s^2}} ,
\end{equation}
and thus, from Eq.~\ref{eq:VDDrho}, it follows that the scale height is given by $H^2 = c_s^2/k$. Defining 
\begin{equation}
\begin{split}
    H_1^2 &= \frac{c_s^2 r^3}{G M_1}, \\
    H_2^2 &= \frac{c_s^2 s^3}{G M_2},
\end{split}
\end{equation}
we finally have
\begin{equation}
    H = \sqrt{\frac{H_1^2 H_2^2}{H_1^2 + H_2^2}}\,.
\end{equation}

In the simulations presented here, the code's artificial viscosity now employs a more accurate estimate of the scale height to maintain a constant Shakura-Sunyaev $\alpha$ throughout the entire simulation, including near the companion. This is the first time a constant $\alpha$ has been properly enforced in simulations of Be binaries.

\subsection{Preliminary tests}\label{sec:prelim}

% \commentACC{O texto desta seção precisa de uma boa revisão, tem muitas repetições e me parece longo demais.}

% \commentACC{Dei uma BOA mudada nesta seção. Ler com atenção.}

%For this work, we wanted to make sure that the modification described in Sect. \ref{sec:psplit} had not disrupted the physical properties of the simulations. In order to test the code, 
The revised code, incorporating the three modifications mentioned above — the updated radius of the secondary's sink, particle splitting, and Shakura-Sunyaev (SS) viscosity parameter $\alpha$ calculated with the corrected $H$ — was tested to ensure that no extraneous effects were introduced in the simulations, and to understand the extent to which the simulations changed with the new implementations.
We calculated a reference model that uses none of the modifications, i.e., the size of the sink of the secondary is the size of its RL, particle splitting is turned off, and the viscosity follows Eq.~\ref{eq:alpha}, but not considering the gravitational effect of the secondary on the scale height of the disk. This is model 1-off, below.
For this model, the Be star and the secondary are as follows: $M_1 = 4.89 \, M_{\odot}$, {$R_{\rm eq} = 6.2 \,R_{\odot}$, $T_{\rm eff} = 13490 \,K$} and $M_2 = 0.97 \,M_{\odot}$, $R_2 = 10^{-6}\, R_{\odot}$. The orbital period is $P_{\mathrm{orb}}$ = 70 days, and the disk viscosity parameter was chosen as $\alpha = 0.4$. \revi{In our simulations, the two stars are sink particles, only serving as gravitational wells for the gas particles that compose the disk. Therefore, the radius of these particles does not directly correspond to the physical radius of the star, rather serving as a boundary inside of which gas particles are accreted. The companion in this case could be interpreted as a compact object or as a stripped star, based on its mass and approximate size, both of which are likely outcomes of previous mass transfer.}
% \addACC{The size of the secondary's sink is written the size of the RL namely $f=XXXXX$ \commentACC{write number}. }

Three additional models were computed using different combinations of parameters, as summarized in Table~\ref{tab:simuls_test}. Model 1-on keeps the same size of the secondary sink but incorporates particle splitting and corrected SS $\alpha$. Model 0.05-off adopts a more realistic size of the secondary sink, making it equal to the physical size of the secondary, but does so without particle splitting and corrected SS $\alpha$. Finally, model 0.05-on represents the full model with all upgrades active.
%The main difference between them is whether the particle splitting option is on or off, and the size of the sink of secondary, in terms of the fraction of its RL radius $f$, as shown in Table \ref{tab:simuls_test}. 

\begin{table}
    \centering
    \caption{Variable parameters for the SPH simulations in Set 1. All other parameters are fixed. $f$ from Eq.~\ref{eq:rlsize}.}
    \label{tab:simuls_test}
    \begin{tabular}{llll}
        \hline
        \hline
         Model & Fraction of $R_L$ ($f$) & Part. splitting \& corrected $\alpha$\\
         \hline
         1-off & 1.0 & Off \\ %twin2
         1-on & 1.0 & On\\ %twin3
         0.05-off & 0.05 & Off\\ %twin
         0.05-on & 0.05 & On\\ %jon's model

         \hline
    \end{tabular}
\end{table}

% Model 1-off uses the ``standard'' version of the SPH code, with none of our modifications active. As such, it is similar to the models presented in past works. 
% such as \citet{panoglou2016}, that characterises the behaviour of Be disks with companions in circular, coplanar orbits. In their discussion of the their results of the temporal evolution of the disk, the models are always compared at the same orbital phase $p$, with $p \in [0,1)$, where $p = 0$ is the apastron, and $p= 0.5$ the periastron. Their simulations have a constant mass injection into the disk. Thus, as the models evolve with time, they reach a point of quasi steady-state, where comparing subsequent cycles at the same phase very little difference can be noted. \citet{panoglou2016} also uses mainly azimuthally averaged quantities in their analysis of their SPH models of binary Be stars. The same quantities for different models are compared for all at the same orbital phase $p$.

% \commentACC{Amanda, muito cuidado com as definicoes. Acima, voce chama o raio da primária de R1, agora eh Req. Escolha uma forma, e use-a de forma consistente no artigo todo, inclusive figuras.}

\begin{figure}
    \centering
    \includegraphics[width=\linewidth]{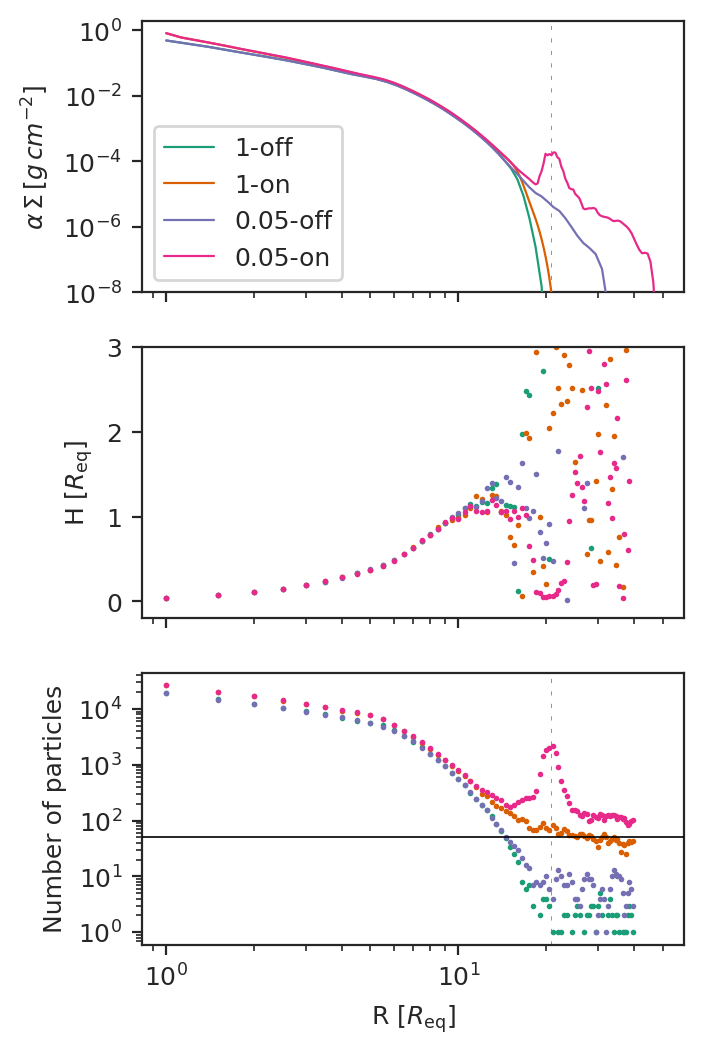}
    \caption{Comparison of averaged surface density and disk height between the models detailed in Table \ref{tab:simuls_test}. First panel shows the density normalized by the viscosity parameter; second panel shows the scale height $H$, and the last panel the number of particles in a given radial extent. The dashed vertical line represents the position of the secondary, which is the same for all simulations. The horizontal line in the bottom panel marks 50 particles.
    % \commentACC{melhorar legenda, descrevendo exatamente o que cada painel mostra}
    % \commentACC{Faltou falar no texto pq dos 50 particulas (linha vertical).}
    }
    \label{fig:prelim}
\end{figure}

In the upper panel of Fig. \ref{fig:prelim} we compare the 
% \commentACC{é a média azimutal? Se sim, definir como é calculada}
azimuthally averaged surface density of all models after 40 completed orbits. Since the effects of particle splitting, corrected SS $\alpha$, and the secondary's sink are only relevant closer the companion, which is located at 20.8\,\req (dashed line in the figure), the density of the inner disk ($\lesssim 10$\,\req) in all models is identical.
%\addACC{The same is true for the disk height shown  in the second panel of Fig.~\ref{fig:prelim}}.
This shows that the modifications have not altered the inner disk and thus our new simulations are comparable to the ones presented in previous works. % \commentACC{isso merece mais destaque, pois confirma que as modificações não tiveram impacto no disco interno, como esperado}

As the particles get closer to the secondary, differences start to show. Particles are accreted (or ``killed'') as soon as they reach a sink particle. In models 0.05-off and 0.05-on, where the sink radius of the secondary is the same as the physical size of the star, particles are allowed to move closer to the companion without being removed from the the simulation. Thus, there is a less sharp cutoff in the density. 
A similar, but much less pronounced, effect is seen for model 1-on, where the resolution of the region close to the secondary is increased by particle spitting.

\revi{There is a density increase around the location of the secondary in models 0.05-on and 0.05-off, as particles can exist in close proximity to the secondary star. In model 0.05-off, however, the total number of particles in the simulation (as seen in the bottom panel of Fig.~\ref{fig:prelim}) prevents this region from being adequately resolved. The use of particle splitting in model 0.05-on circumvent this issue, allowing for a better resolution, showing a much larger accumulation of matter in this region.} On the other hand, in model 1-on, there are a sufficient number of particles at large radii, but the substantial size of the secondary's sink effectively prevents matter from being captured in its gravitational well. Consequently, only model 0.05-on, which incorporates all modifications to the code, enables the resolution of the structure around the secondary. This is a significant finding that will be thoroughly explored in this paper.
% \commentACC{acho que ficaria muito legal colocar um novo painel na figura 1 mostrando o número de partículas como função do raio. Podemos colocar uma barra de erro poisonica para indicar as regiões bem resulvidas e mal resolvidas}
% \rmACC{Comparing the density structure of models 1-on and 0.05-on it also becomes clear that in order to see the matter around the secondary, having a sink radius corresponding to the actual radius of the companion is necessary. }

The middle panel of Fig.~\ref{fig:prelim} shows the disk scale height, $H$. It was calculated by fitting a Gaussian to the volume density $\rho(r,z)$ along the vertical direction for 1 \req slices along $r$. As expected, the models all behave identically in the inner disk, again assuring that model not unduly disrupted by the modifications. Model 0.05-on has the highest resolution of all models, given the particle splitting, which allowed its height to be measured with greater confidence up to 30 \req, a region which did not exist, for all intents and purposes, in the previous works in Be literature with this SPH code. As shown in the bottom panel of Fig.~\ref{fig:prelim}, models without particle splitting have fewer than 50 particles in the radial bands after about 12 \req, making the disk structure unresolved and the scale height and density determination completely unreliable.

In conclusion, our tests confirm that the new features integrated into the code do not hinder us from obtaining results consistent with those of the old code for areas where no differences between them were anticipated. Additionally, we demonstrate that we are now capable of investigating previously obscured structures of the system, including the area around the companion and beyond. \revi{Table~\ref{tab:upgrades} offers an overview of our upgrades to the code used in this work.}

\begin{table*}
    \centering
    \caption{Overview of the upgrades to the SPH code. }
    \begin{tabular}{c c c}
    \hline
    \hline
          & Previous works & This work \\
         \hline
         Secondary's sink & Size of its RL &  Size of the star\\
         Particle splitting & Not implemented & Implemented\\
         Viscosity prescription & Not accounting for the secondary & Accounting for the secondary\\
         \hline
    \end{tabular}
    \label{tab:upgrades}
\end{table*}

% IS THIS ENOUGH OF A TEST?
% \commentACC{Acho que sim, foi feito tudo o que pensamos em fazer}

\section{Models}\label{sec:models}

%\commentACC{Normalmente nos papers nossos a gente fala em mass injection rate, não ejection. Injection no disco. Sugiro manter essa nomenclatura}
Using the updated SPH code, we ran a series of simulations with a focus on the previously unresolved regions of the system. \revi{As most Be stars in binaries are early-type Be stars, we based this set of models on the well-known Be binary system $\pi$\,Aqr \revi{(B1III-IVe)}. This system was also believed to host more massive companion than usual for Be binaries, consequently leading to stronger effects on the disk dynamics, which is our object of study. We use the values determined by \citet{bjorkman2002} for the parameters of the Be star and keep the radius of the secondary fixed at 1 $R_{\odot}$}. The fixed parameters in our simulations are shown in Table~\ref{tab:params}.
% They are based on the well-known Be binary system $\pi$ Aqr. We use the values determined by \citet{bjorkman2002} for the parameters of the Be star and keep the radius of the secondary fixed at 1 $R_{\odot}$. 
Our full set of simulations is comprised of 11 models where the modifications to the code (Sect. ~\ref{sec:sphcode}) are always on, and only the orbital period, mass of the secondary (thus the mass ratio $q$)\footnote{Most simulations have a mass ratio of 0.16, based on \citet{bjorkman2002}. We note, however, that since the start of our work, the mass of the companion of $\pi$ Aqr has been put to question by \citet{tsujimoto2023}, who finds $M_2 < 1.4\,~M_{\odot}$ in contrast with \citet{bjorkman2002}'s $M_2 = 2.06\,~M_{\odot}$.}, and viscosity parameter $\alpha$ are allowed to vary, as described in Table \ref{tab:inner}. 
% \commentACC{Amanda os nomes do texto abaixo e da tabela 3 nao estao batendo, revisar com cuidado}
For example, model 30-0.1-0.16 has a period of 30 days and $\alpha = 0.1$ and $M_2 = 2.06~M_{\odot}$ (the mass ratio $q = 0.16$). Models 30-0.5-0.16 and 30-1.0-0.16 have the same period and $M_2$, but $\alpha = 0.5$ and 1.0, respectively. The models designated by 50 and 84 are similar to the three just described, but with different orbital periods. We note that $\pi$~Aqr's orbital period is 84.1 days.
Finally, models 30-1.0-0.33 and 30-1.0-0.50 are the same as model 30-1.0-0.16
% $P_{\mathrm{orb}}$ = 30 days and $\alpha = 1.0$, 
but with a different secondary mass (4.30~$M_{\odot}$ for $q = 0.33$, and 6.45~$M_{\odot}$ for $q = 0.5$). \revi{The companion masses were chosen to represent the higher mass end of stripped stars and He stars, the likely results of a previous mass transfer phase with the Be star, and including MS stars as well. Using more massive companions also maximizes its gravitational effects on the Be disk, aiding our characterization of the dynamics of these systems. }

In the simulations presented here the mass injection rate ($\dot{M}_{\mathrm{ij}}$) is kept fixed \revi{at the value of $10^{-8}\,M_{\odot}\,\mathrm{yr}^{-1}$, which has been shown to produce disk densities in the typical range of Be star disks \citep{panoglou2016, vieira2017}. Although this choice is physically motivated, it is should be noted that the mass of the gas particles is in truth an arbitrary value, since the disk is not self-gravitating. Therefore, the masses scale with the number of particles and on their status as children of split particles. } 
%\commentACC{Sugestão}
% \addACC{This, together with the particle mass of XXX, has been shown to produce disk densities in the typical range of Be star disks \citep{panoglou2016, vieira2017}.
% However, it is should be noted that, since the disk is not self-gravitating, both the mass injection rate and particle mass are somewhat arbitrary parameters that can be toggle within reasonable ranges without affecting the geometry or dynamics of the disk.
% }
%\commentACC{Esta parte do injection zone não foi definida acima quando descreve o código.}
Most of the particles fall back into the central star due to the natural loss of angular momentum caused by interactions with the matter already in orbit and only about 0.1$\%$ of the particles manage to maintain orbit and thus are responsible for building the disk up. %This means that the mass decretion rate, i.e., the rate at which mass is effectively lost from the Be star, is a number of the order of $10^{-10}$ to $10^{-9}\,M_{\odot}\,\mathrm{yr}^{-1}$, depending on the model.

%The mass \textit{injection} rate, then, is about $\dot{M}_{\mathrm{inj}} = 10^{-9} M_{\odot}/\mathrm{year}$. 

The disk is isothermal in our simulations, with an electron temperature fixed at 60$\%$ of the effective temperature of the Be star \citep{carciofi2006}.
As shown in Sect.~\ref{sec:diskphysics}, the density of a Be disk in the steady-state VDD formulation falls off with radius as $\rho \propto r^{-3.5}$ for a thin, isothermal disk. However, the temperature structure of real Be disks is decidedly not isothermal. \citet{carciofi2006, carciofi2008}, and more recently \citet{suffak2023} used the 3D \revi{non-local thermodynamic equilibrium (NLTE)} Monte Carlo radiation transfer code \textsc{hdust} to simulate the temperature structure of a VDD with a fixed density structure and a single star as the radiation source. They find that the temperature {at the midplane} is high close to the Be star, but drops for larger radii, before coming up again and reaching a plateau \citep[see, e.g., Fig.~3 of][]{carciofi2008}. 
% \commentACC{Citar aqui o artigo recente do Mark sobre a estrutura de termica de discos tilted. Ele também analisa os resultados para discos coplanares, extendendo os resultados do meu artigo de 2006}
This property of the disk affects its structure: {the decrease in temperature in the midplane leads to a decrease in the scale height $H$ \citep[Eq.~\ref{eq:scaleh}, see also][]{carciofi2008}, which in turn increases the density in this region and decreases in the density of the upper layers.}

It is therefore quite clear that non-isothermal effects influence the disk structure of Be stars and their observables. For the discussion presented here we assume isothermality for simplification. Developing a non-isothermal version of the SPH code is feasible, such as by integrating SPH with \textsc{hdust}, but it constitutes a substantial undertaking that falls beyond the scope of this paper.

% \commentACC{Eu tiraria esta parte. Para explorar os efeitos não-isotérmicos nos modelos, precisariamos de um modelo conjunto SPH+HDUST que não existe. Basta dizer, na minha opinião, que o SPH considera que o gás é isotérmico, e isso é uma aproximação - que alias é feita por todo mundo}
% It is therefore quite clear that non-isothermal effects influence the disk structure of Be stars and their observables. For the discussion presented here, however, it was determined that exploring different temperature structures in our SPH simulations would be overhasty, since although they would have an impact, it would not as substantial and consistent as the ones caused by different binary periods, mass ratios, and disk viscosities. ESSE PARÁGRAFO PODE TER QUE MUDAR SE FORMOS EXPLORAR A TEMPERATURA COM O HDUST!

  % \commentACC{Nõa acho necessário seguir por este caminho, da consistência. Afinal, $q$ é um parâmetro livre. Eu diria simplesmente que por conveniência escolheemos a massa de 12,9 de pi Aqr, que é representativa de uma estrela B1, que corresponde ao pico da distribuição de estrelas Be na Galáxia -- ou seja, é uma Be típica}

%, and orbital period (which has been confirmed unambiguously by subsequent works such as \citealt{zharikov2013}). VALORES VALORES... TABELA? 
% A low value for the viscosity of $\alpha = 0.1$ was chosen, and the period of the system was set to 30 days. Models with the system's true period, 84.1 days, were also computed and are presented in Sect. xx.

\begin{table}
    \centering
    \caption{Fixed parameters of the SPH simulations. }%\commentACC{de novo, cuidado com os simbolos. Voce usa as vezes Teff as vezes T1. Escolher um e fixar sempre}
    % \commentACC{Não vejo necessidad desta tabela ser de pagina inteira. Penso que ficaria melhor seguir o modelo da tabela 1}
    % }
    \begin{tabular}{c c c c c c c c}
    \hline
    \hline
        $M_{1}$ &  \req & $T_{\rm{eff}}$ & $R_{2}$ & Particle splitting\\
        $[M_{\odot}]$ & $[R_{\odot}]$ & [K] & $[R_{\odot}]$ &  & \\
        \hline
        12.9 & 5.5 & 26000 & 1.0 & On \\
        \hline
    \end{tabular}
    \label{tab:params}
\end{table}

\subsection{Overview of the system}
\label{sec:overview}

% \commentACC{mudei bastante aqui, ler com cuidado}

\begin{figure*}
    \centering
    \includegraphics[scale=1.0]{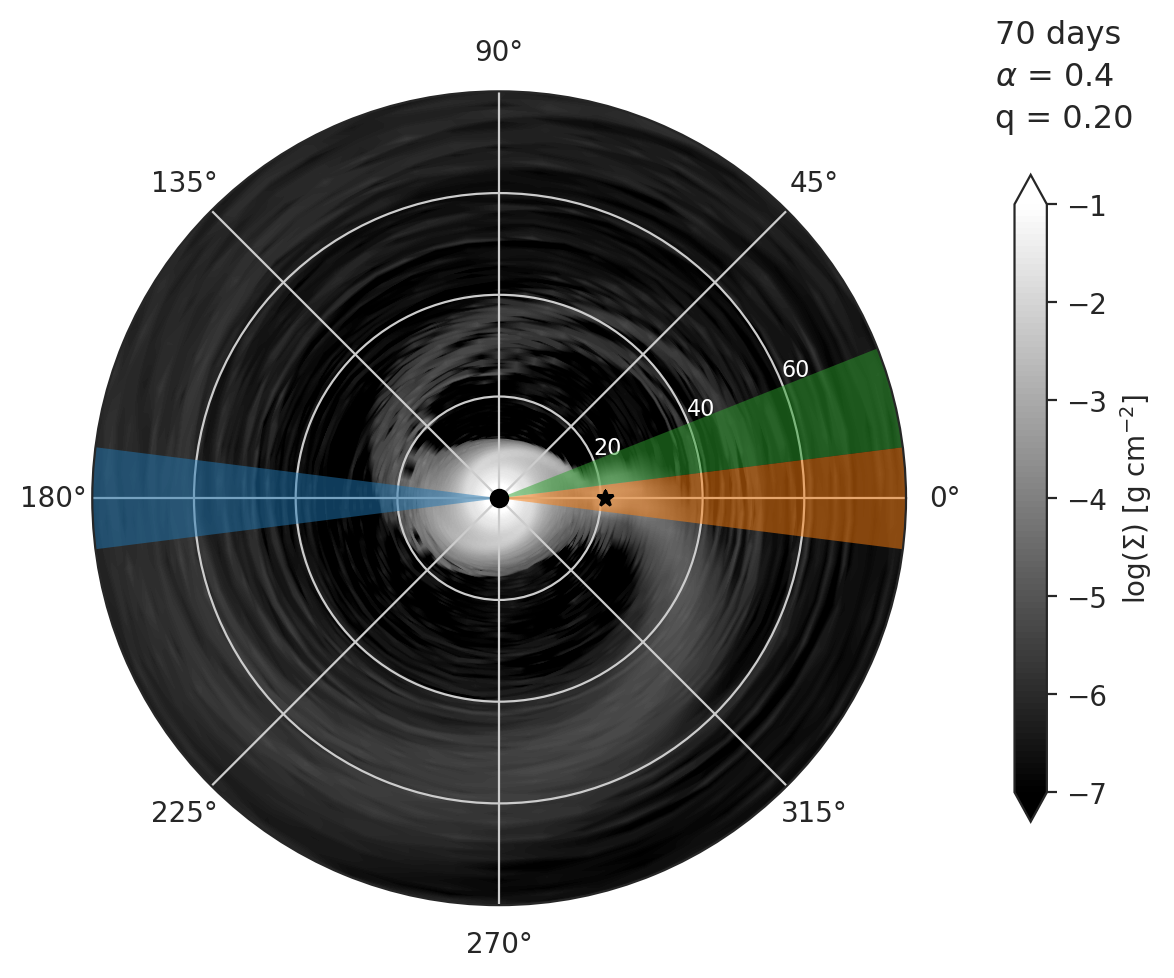}
    \caption{
    % \commentACC{A figura original do Jon tinha um contraste melhor, pq será? Era mais facil ver os 2 bracos externos. Sugestão: tentar tirar (igualar a zero) as regiões com densidade menor que um certo valor, talvez, -7 ou -6. Afinal, não faz sentido explorar 10 ordens de maginute em densidade como faz a figura. Outra: limitar o raio a 80 raios externos. Não dá para ver nada no ultimo anel da figura.}
    Surface density map of simulation 0.05-on. The overall density structure is qualitatively similar for all simulations calculated for this work, as sketched in Fig. \ref{fig:esquema}. \revi{The black circle denotes the Be star, and the black star, the companion.} The white circles and numbers mark the radial distance in \req. The three colored bands correspond to the main regions of interest: the wedge that contains the secondary (yellow), the wedge opposite to the secondary (blue), and a wedge that includes the bridge (green).}
    \label{fig:mapao}
\end{figure*}

\begin{figure}
    \centering
    \includegraphics[width=\columnwidth]{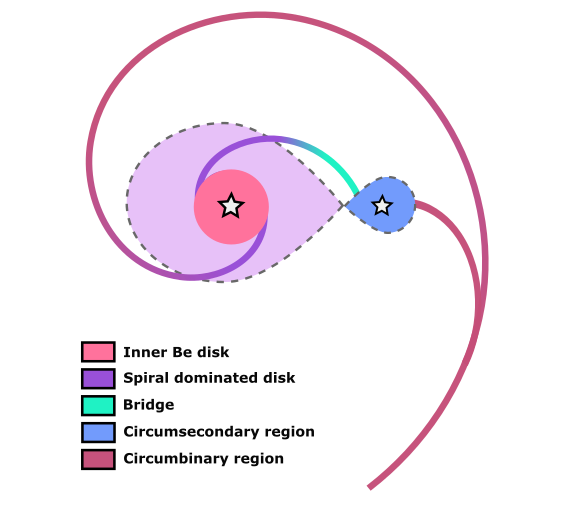}
    \caption{Sketch view of the main regions of the binary Be system. The five main regions are the inner Be disk (pink), the spiral dominated disk (purple), the bridge (teal), the circumsecondary region (blue) and the circumbinary region (mauve). The main characteristics of each region are detailed in the text.}
    \label{fig:esquema}
\end{figure}

% \commentACC{Acho que este parágrafo deveria começar mostrando a Fig. 2 em toda sua gloria -- largura de página, por exemplo. Seria muito legal marcar com cores ou linhas as 5 regiões que voce descreve abaixo}
The most significant aspect of our modifications to the code is that the resolution in the outer disk is much improved when compared to previous works (Sect.~\ref{sec:prelim}). Structures that were previously absent can now be seen clearly in the simulations, as shown in Fig.~\ref{fig:mapao}. In previous works with this SPH code, only the region inside 20\,\req in Fig.~\ref{fig:mapao} was explored, while our simulations cover a region 4 times larger. 
% \commentACC{Temos que valorizar mais a figura aqui. Uma forma de fazer isso é voce mostrar na figura o local aproximado que era explorado pelas simulacoes anteriores (tipo aquela figura de talk que fez) e indicar que nossas simulacoes cobrem uma area XXX vexes maior, tipicamente}
Thus, we define five distinct regions of interest and analyze them separately, as sketched in Fig.~\ref{fig:esquema}: the inner Be disk, the spiral dominated disk, the bridge, the circumsecondary region, and the circumbinary region. 

The inner Be disk is the region least perturbed by the secondary
and typically spans a handful of stellar radii in extent.
The spiral dominated disk comprises the region from the edge of the inner Be disk to the
region that previous studies refer to as the ``truncation radius''. In actuality, our simulations show that calling it a truncation is misleading, since it refers in fact to a dynamically complex region that stretches for several stellar radii; \fix{\citet{suffak2022} uses the more appropriate term \textit{transition radius}}. The denser, leading arm extends into the RL of the companion, dumping matter from the Be disk into it. We call this the bridge, as it connects the Be disk to the secondary. Matter accumulates around the secondary forming what we name circumsecondary region. Some of the particles that form this region will be accreted, but part will have enough energy to exit the RL behind the companion. These escaping particles meet up with the other spiral arm, forming one big spiral that envelops the entire system, which we call circumbinary region. Neither the circumsecondary or circumbinary regions were present in any of the previous works that used this SPH code. Each of these regions is explored in the following sections.

% TRABALHANDO NUMA FIGURA ESQUEMA PRA MOSTRAR AS DIREÇÕES DA CUNHA E AS REGIÕES

% STEP STACKING
Since our simulations have a steady mass injection, once the disk is fully formed and the accretion onto the secondary also stabilizes, the whole system reaches a regime of quasi steady state \citep{panoglou2016}.
% \commentACC{citar Despina, ela descreve bem esse estado de quasi steady-state}
This means that between subsequent timesteps there is no change in the physical quantities other than that created by the SPH noise and, of course, the rotation of the system. Thus, if the rotation of these steady-state timesteps is frozen, i.e., if we fix the two stars along the $x$-$y$ axis of the corotating system, the particles present in each of these timesteps can be combined into one single ``stacked'' timestep. Since our main focus is on the structures formed after the steady state is reached, this method allows us to artificially increase the resolution of the whole simulation.

% WEDGES
%\commentACC{Cuidado aqui, apenas a Despina usou essas quantidades, e ainda assim não apenas elas.}
%Most previous works on Be binaries SPH simulations choose to analyse the physical quantities of the simulation by doing azimuthal averages. 
As Fig.~\ref{fig:mapao} shows, the system is not axissymmetric, changing considerably in shape and density along different azimuthal directions. In order to quantify and characterise the different regions of the system, we choose to explore its physical quantities (e.g., velocity, density, etc.) along %these different lines of sight by defining 
wedge-shaped regions of interest fixed at certain azimuthal angles.
Within a given wedge, we compute the quantities at different radial distances and heights (i.e., distance above the equatorial plane).
%, composed of cells in radial, azimuthal and vertical directions. We then calculate the average quantities in each cell using the particles in those regions. 
%\commentACC{Mais detalhes são necessários sobre como as quantidades medias são calculadas} 
The average quantities are defined as the mass-weighted mean value in a cell, which allows us to account for the different masses of the particles and ignore empty cells in the grid.
%\rmACC{Using this method, we not only obtain a more complete view of the system, but are also able to study how the observables vary depending on the azimuthal angle and height.} 
For the purposes of this work, certain azimuthal directions (given by the angle $\phi$ in Fig.~\ref{fig:mapao}) have more distinct features and are more interesting to be highlighted. {These are the direction of the secondary, by definition $\phi = 0$ (in yellow), the direction opposite to the secondary, $\phi = 180^{\circ}$ (in blue) and the direction containing the bridge between the Be disk and the companion, $\phi = 20^{\circ}$ (in green).} 
%\commentACC{???} era coisa velha q me escapou
Fig.~\ref{fig:mapao} highlights these three wedges of interest. % in 

\subsection{Inner Be disk}\label{sec:inner_be_disk}

% \ACR{Mudei a ordem dos parágrafos aqui}

% A FAZER AQUI; grafico de sigma vs. phi para diferentes r (dentro do disco interno). Há variação com phi, em que nível? NO VARIATION!

% \commentACC{reler com cuidado esta secao. Há alguma repetição do texto, acho que dá para enxugar uns 30 - 40\%. }

% \commentACC{Amanda, acho que esta secao precisava de algum trabalho. Para facilicar fui cortando os blocos de texto com o rmACC e rearranjando com o addACC}

% \commentACC{Em alguns momentos voce antecipava o spiral dominated disk. Eu achei que isto deixava o texto confuso então tirei essas partes, com exceção de uma mencao ao final}

% \rmACC{In our SPH code, mass is lost by the star through the equator in all azimuthal directions.
% For the inner Be disk, the mass flux is then the same as the mass injection rate ($\dot{M}_{\rm inj}$), a fraction of the mass ejected ($\dot{M}_{\rm ej}$) by the central star. In all models, only 0.1$\%$ of the ejected mass effectively goes on to build up the Be disk; the rest is immediately reacreted by the Be star.}
% \commentACC{confusao de termos acima. Mdot no SPH é o mass injection rate. Mas flux é o fluxo que passa em determinado ponto. Em steady state, mass flux = mass decretion rate. Agora, eu acabei fazendo já esse comentário acima, então acho que não precisa repetir isso aqui.
% }

\begin{figure}
    \centering
    \includegraphics[scale=0.5]{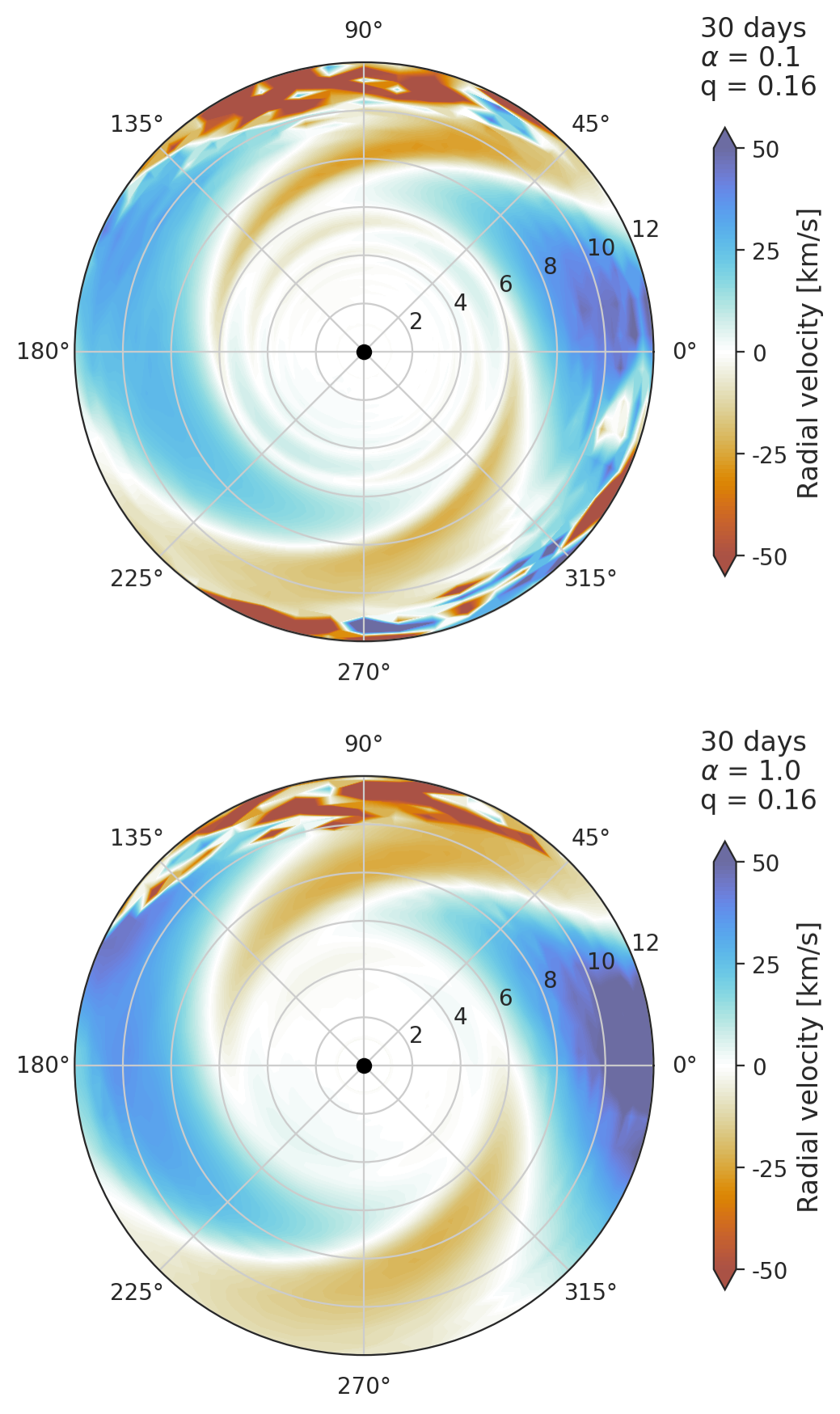}
    \caption{Radial velocity maps for models 30-0.1-0.16 and 30-1.0-0.16. The black dot in the center represents the central Be star. The double spiral structure formed by the elongation of the orbits of the particles is visible in both maps. The overdensity occurs in the the apastron of the orbits of the particles, where the particles slow down. 
    % \commentACC{Amanda, talvez a escolha de cores desta figura não esteja boa, pois a gente nao consegue ver claramente o zero}
    }
    \label{fig:velcompare}
\end{figure}

In the SPH simulation, mass is lost by the star through the equator at an equal rate in all azimuthal directions. Therefore, one expects that the disk should maintain an axisymmetric configuration near the star and display the signs of the secondary's action only at greater distances. This is exactly what is observed in Fig.~\ref{fig:velcompare}, that compares the radial velocities within the first 12 stellar radii of the disk (as per the zoomed in view in this Figure compared with Fig.~\ref{fig:mapao}).
The figure contrasts models 30-0.1-0.16 and 30-1.0-0.16 and confirms that within roughly 3 stellar radii, model 30-0.1-0.16 has a nearly circular configuration. This is also observed in the surface density and azimuthal velocity. We call this region the inner Be disk (pink region in Fig.~\ref{fig:esquema}).

To make a quantitative assessment of the size of the inner disk, we define that the transition to the spiral dominated disk (see the purple region in Fig.~\ref{fig:esquema} and Sect.~\ref{sec:spiral}) happens when the azimuthal density variation caused by the presence of the companion exceeds 5$\%$. Since the density waves are caused by the changes in the orbits of the disk particles due to tidal interaction with the companion, this shift can be seen in the radial velocity maps of the disk. Fig.~\ref{fig:velcompare} shows that the spiral waves are noticeable already at about 4 \,\req for the lower $\alpha$ simulation, while they only become relevant at around 6\,\req for higher $\alpha$.
%for higher $\alpha$ shows that the spiral waves are noticeable already at about 4 \,\req for the lower $\alpha$ simulation, while they only become relevant at around 6\,\req for higher $\alpha$.

The inner disk sizes for all models are listed in Table~ \ref{tab:inner}, and clearly depend on the period and viscosity more significantly than on the mass ratio, although its effect is not completely negligible. Given the orbital period and the stellar masses, the separation of the two stars is

\begin{equation}
    a = \left[\frac{G (M_{1} + M_{2}) P^2}{4 \pi^2}\right]^{1/3}\,.
\end{equation}

% \commentACC{faltava o G acima}
\noindent It follows that for larger periods, where the binary separation ($a$) is larger, the larger the size of the inner disk. This can be seen comparing models with different periods, such as, for instance, 30-1.0-0.16, 50-1.0-0.16 and 84-1.0-0.16. {Increasing the viscosity also leads to a larger inner disk, as a more viscous disk grows faster and is more resilient against the tidal torque exerted by the companion. 
Quantitatively, we can say that for $\alpha = 0.1$, the size of the inner disk approximately is 20\% of the binary separation $a$; for $\alpha = 0.5$, it is closer to 28\%, and for $\alpha = 1.0$, around 30\%. The effect of the mass ratio is seen when we consider that $a$ also depends on the mass of the two stars: the size of the inner disk is the same between models 30-1.0-0.33 and 30-1.0-050, as the gravitational effects of the companion are also stronger, balancing out the larger binary separation.}

In the inner disk, the dominant forces acting on a gas particle are the gravity of the Be star and the viscous force.
Even so, the effects of the companion in this region are not completely negligible. The tidal torque removes part of the angular momentum (AM) of the gas, leading to an accumulation of matter \citep[this is the aforementioned accumulation effect of][]{panoglou2016}. The net result is a decrease of the surface density radial exponent from the canonical steady-state VDD value of $m=2$ (Eq.~\ref{eq:VDDsigma}). The extent of this change in $m$ is very model dependent, as seen in Table~\ref{tab:inner}, where we estimated the value of $m$ for each model by fitting a power-law to the azimuthal mean of the surface density of the inner disk. The role of each parameter in shaping $m$ is explored below.

\begin{figure*}
    \centering
    \includegraphics[width=\textwidth]{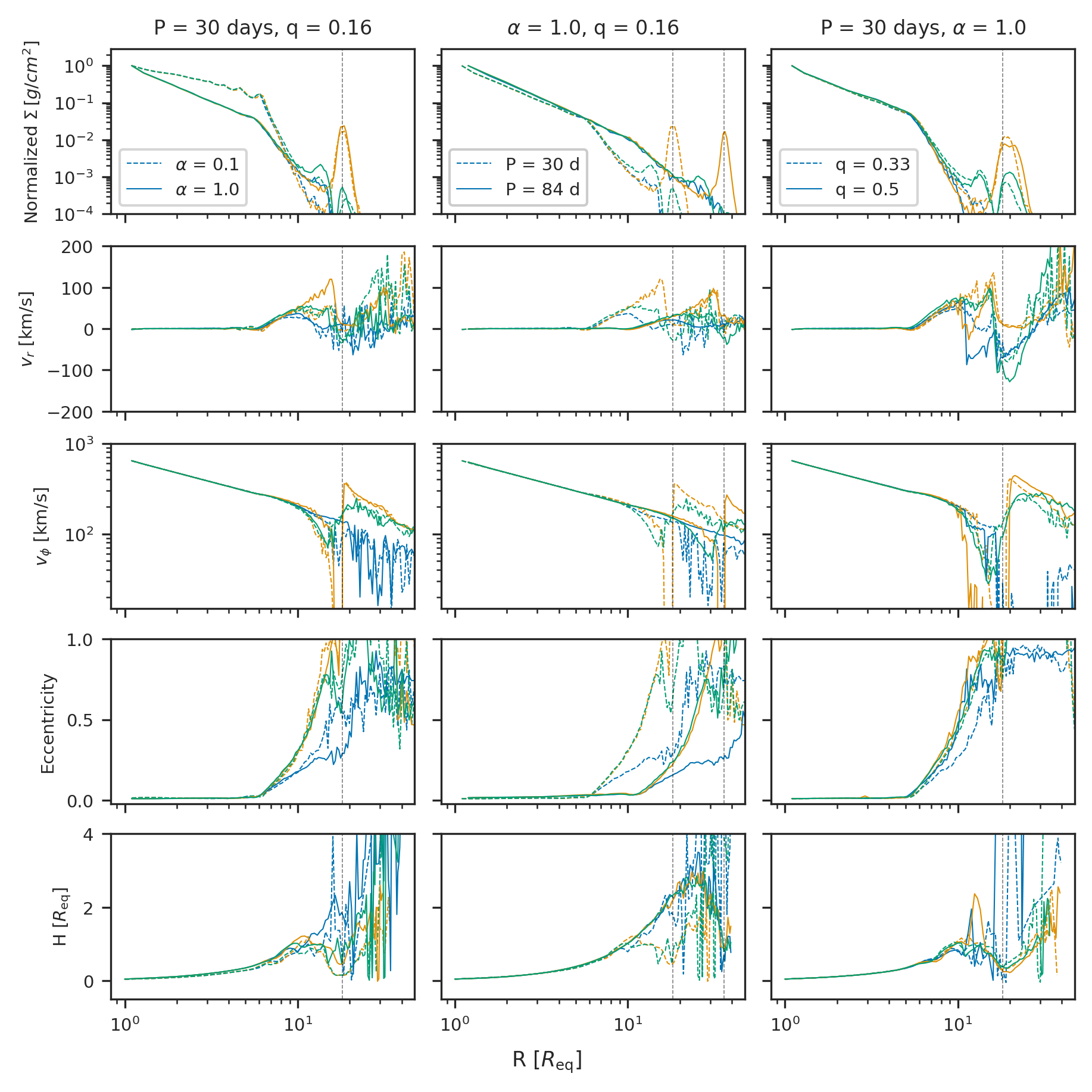}
    \caption{Surface density, radial velocity, azimuthal velocity, eccentricity of the particles' orbits, and scale height varying with radius for models 30-1.0-0.16 and 30-0.1-0.16 in the leftmost panels, 30-1.0-0.16 and 84-1.0-0.16 for the middle panels, and 30-1.0-0.33 and 30-1.0-0.50 for the rightmost panels. The colors of the lines indicate the wedges along which the quantities are calculated, following Fig. \ref{fig:mapao}: yellow is the wedge pointing to the secondary, blue is the wedge in the opposite direction, and green includes part of the stronger spiral arm. The dotted gray lines show the position of the secondary for the models compared in each column.}
    \label{fig:grid}
\end{figure*}

Figure~\ref{fig:grid} compares the surface density, radial velocity, azimuthal velocity, eccentricity of the particles' orbits, and scale height for several of our models. The accumulation effect is clear when we compare models 30-0.1-0.16 and 30-1.0-0.16, in the top-left panel of Fig.~\ref{fig:grid}. Model 30-1.0-0.16 follows a density drop-off (Eq.~\ref{eq:VDDsigma}) with $m \simeq 2.0$, as expected from an isolated Be star. Model 30-0.1-0.16, on the other hand, has a much denser inner disk, in addition to being radially smaller ($3.7\pm0.2$ \req, $m = 0.89$). For the same $\alpha$ and $q$, the larger the inner disk, the smaller the accumulation of matter, thus larger the value for $m$. Larger mass ratios, such as for models 30-1.0-0.33 and 30-1.0-0.50 lead to larger $a$, but comparatively smaller $m$ (compared to model 30-1.0-0.16). The disk viscosity also has a significant impact on the accumulation effect. The models indicate that increasing $\alpha$ consistently raises $m$ (i.e., reduces the accumulation effect), \fix{which also agrees with previous determinations of $m$ by \citet{cyr2017}}. This is simply a consequence of the balance between the viscous torque and the tidal torque of the secondary in this region: the more viscous disks are less subject to the tidal effects of the companion.

Nevertheless, the inner Be disk is a stable region when compared to the rest of the system. The radial and azimuthal velocities ($v_r$ and \vp) of the particles follow the expected for a quasi-Keplerian motion: $v_r$ increases only slowly and monotonically while \vp follows a power-law with $r^{-0.5}$, and the eccentricities of the orbits of the particles are also very low for all models. {The scale height of the disk also remains remarkably consistent in this region, only deviating in the regime of the spiral dominated disk.}

% \commentACC{Em vista das sugestões acima, penso que este parágrafo possa ser removido.}
% \rmACC{
% The massive impact of the viscosity parameter in behaviour of the inner disk can be understood when we consider that viscosity if what controls the growth rate of the disk. The viscous diffusion timescale is given by}
% \begin{equation}
%     t_{\rm diff} = \frac{V_{\rm crit}}{\alpha c_s^2} \, r^{1/2} .
% \end{equation}\label{eq:visctimescale}
% \rmACC{
% Thus, the timescale for the build up of the disk is lower for larger viscosity (larger $\alpha$), or, simply put, more viscous disks grow faster. The tug-of-war between the viscous torque and the resonant torque of the companion, which steals AM from the particles and stops them from moving to higher orbits, is responsible for both the excitation of the density waves and the accumulation effect. Since the more viscous the disk, the higher its viscous torque, both the waves and accumulation effect are dimmed in comparison with low viscosity disks. }

 % The accumulation effect is a consequence of the tug-of-war between the viscous torque and the resonant torque of the companion, which steals AM from the particles and stops them from moving to higher orbits. When $\alpha$ is larger, the resonant torque has a smaller effect, and the accumulation of matter in the inner Be disk is not as significant. On the other hand, if $\alpha$ is low, the accumulation effect is more effective. As shown in Fig. \ref{fig:chakra}, the radial density exponent $m$ of Eq. \ref{eq:VDDsigma} goes as low as $m = 1$ for simulations with $\alpha = 0.1$.

% \begin{table*}
%     \centering
%     \caption{Size of the inner disk region for each model, and the winding parameter $\gamma$ for the two spiral arms (leading - trailing, respectively).}
%     \label{tab:inner}
%     \begin{tabular}{lccc}
%         \hline
%          Model name & Size of inner disk [$R_{\rm{eq}}$] & Winding param. ($\gamma$) & $m$ (Eq. \ref{eq:VDDsigma})\\
%          \hline
%          30-0.1-0.16 & 3.7 & 0.50 - 0.50 & 0.93\\
%          30-0.5-0.16 &  5.3 & 0.32 - 0.39 & 1.76\\
%          30-1.0-0.16 & 5.3 & 0.21 - 0.28 & 2.01\\
%          \hline
%          50-0.1-0.16 & 4.9 & 0.33 - 0.35 & 1.07\\
%          50-0.5-0.16 & 7.1 & 0.21 - 0.27 & 1.72\\
%          50-1.0-0.16 & 7.1 & 0.17 - 0.20 & 1.94\\
%          \hline
%          84-0.1-0.16 & 6.6 & 0.23 - 0.24 & 1.20\\
%          84-0.5-0.16 & 9.3 & 0.14 - 0.18 & 1.67\\
%          84-1.0-0.16 & 9.3 & 0.13 - 0.16 & 1.82\\
%          % \hline
%          % 100-01-16 & 8.1 & 0.20 - 0.21\\
%          % 100-05-16 & 10.7 & 0.12 - 0.16\\
%          % 100-1-16 & 11.4 & 0.13 - 0.13\\
%          \hline
%          30-1.0-0.33 & 5.1 & 0.27 - 0.33 & 1.92\\
%          30-1.0-0.50 & 4.9 & 0.34 - 0.38 & 1.82\\
%          \hline
%     \end{tabular}
% \end{table*}

\begin{table*}
    \centering
    \caption{Free parameters for the complete set of SPH simulations (all other parameters are fixed). 
    The last five columns show the binary separation $a$, the estimated size of the inner disk, the winding parameter $\gamma$ for the two spiral arms (leading - trailing, respectively), the estimated exponent of the mean surface density profile of the inner disk, and the number of orbits necessary for the full build-up of the Be disk.}
    % \commentACC{Seria muto util ter o $a$ aqui}\answerACR{hm, não sei, pq é o mesmo valor pros modelos com o mesmo período e mesmo q} \commentACC{Sim, as este valor não está listado em lugar nenhum. Alem disso, comparar o tamanho do disco interno com $a$, para um mesmo alpha, é interessante}
    % }
    \label{tab:inner}
    \begin{tabular}{lcccccccc}
        \hline
         Model name & P$_{\rm orb}$ & $\alpha$ & $q$ & $a$ & Size of inner disk & $\gamma$ (Eq.~\ref{eq:winding})  & $m$ (Eq.~\ref{eq:VDDsigma}) & Be disk build-up\\
         & [days] &  & &[$R_{\rm{eq}}$] & [$R_{\rm{eq}}$] &  &  & time [orbits]\\
         \hline
         30-0.1-0.16 & 30 & 0.1 & 0.16 & 18.20 & 3.7$\pm$0.2 & 0.50 - 0.50 & 0.89 & 21.2\\
         30-0.5-0.16 & 30 & 0.5 & 0.16 &  18.20 &5.3$\pm$0.2 & 0.32 - 0.39 & 1.76 & 10.4\\
         30-1.0-0.16 & 30 & 1.0 & 0.16 &  18.20 &5.3$\pm$0.2 & 0.21 - 0.28 & 2.01 & 9.3\\
         \hline
         50-0.1-0.16 & 50 & 0.1 & 0.16 & 25.59 & 5.2$\pm$0.3 & 0.33 - 0.35 & 1.07 & 16.7\\
         50-0.5-0.16 & 50 & 0.5 & 0.16 & 25.59 & 7.1$\pm$0.3 & 0.21 - 0.27 & 1.74 & 13.1\\
         50-1.0-0.16 & 50 & 1.0 & 0.16 & 25.59 & 7.4$\pm$0.3 & 0.17 - 0.20 & 1.94 & 10.1\\
         \hline
         84-0.1-0.16 & 84.1 & 0.1 & 0.16 & 36.19 & 7.0$\pm$0.4 & 0.23 - 0.24 & 1.21 & 13.8\\
         84-0.5-0.16 & 84.1 & 0.5 & 0.16 & 36.19 & 9.8$\pm$0.4 & 0.14 - 0.18 & 1.70 & 11.1\\
         84-1.0-0.16 & 84.1 & 1.0 & 0.16 & 36.19 & 9.8$\pm$0.4 & 0.13 - 0.16 & 1.83 & 7.4\\
         % \hline
         % 100-01-16 & 8.1 & 0.20 - 0.21\\
         % 100-05-16 & 10.7 & 0.12 - 0.16\\
         % 100-1-16 & 11.4 & 0.13 - 0.13\\
         \hline
         30-1.0-0.33 & 30 & 1.0 & 0.33 & 19.07 &5.3$\pm$0.2 & 0.24 - 0.30 & 1.93 & 3.2\\
         30-1.0-0.50 & 30 & 1.0 & 0.50 & 19.83 &5.3$\pm$0.2 & 0.28 - 0.30 & 1.87 & 2.7\\
         \hline
    \end{tabular}
\end{table*}

% Fig.~\ref{fig:zoom_rainbow} compares $v_{r}$ and \vp of the particles along different azimuthal directions. For the inner disk, there is only minor deviations from the Keplerian motion.

% The inner Be disk, by our definition, ends when the spiral arms become the most significant characteristic of the density and velocity profiles of the disk. This happens where the isodensity contours shown in Fig. \ref{fig:isocontours} become elliptical and in Fig. \ref{fig:rainbow} where the orbits of the particles begin to greatly deviate from circular, and \vp begins to deviate from Keplerian.

\subsubsection{Observational expectations}

Confirming Be stars as binaries is not straightforward. The standard method for binary detection is spectroscopy, where the radial velocity shifts of the spectral lines can trace orbital motion. In most (all?) cases, Be stars are not found in systems with Main Sequence companions, 
% \commentACC{Comentário do Dietrich em outro artigo: Isto é uma citação incompleta do trabalho da Julia, pois ela focou sua busca apenas para estrelas mas recentes que 1.5.}
but rather with evolved stars (\citealt{bodensteiner2020} found no early-type Be+MS binaries in their sample of 287 Galactic Be stars). Thus, as the Be star is the more massive and luminous component of the system, the combined spectra is heavily dominated by the rotationally broadened Be star lines and emission lines from the Be disk, making radial velocity measurements in the visible a challenge. Therefore we look for signs of Be disk disruption in observational data that can point towards hidden close objects. Here we explore which observational characteristics of the inner Be disk may indicate binarity.

% \commentACC{Hmm, não sei se é uma boa o que esta escrito. Eu falaria simplesmente que detectar binarias Be é desafiador pois em geral o mass ratio é alto e as linhas fotosfericas são muito alargadas e frequentemente contaminadas por emissão, fazendo com que medidas confiáveis de velocidade radial sejam difíceis}
% Confirming Be stars as binaries in the absence of X-ray emission is not straightforward. It is therefore in our interest to look for signs of Be disk disruption in observational data that can point towards hidden close objects. 

The inner Be disk, excluding the Be star itself, is the densest region of the system. It may dominate emission in the short wavelength continuum (from the visible up to the the near-infrared), and frequently does so for longer wavelenghts. The disk is also responsible for most of the polarized flux.
\citep{carciofi2011}. As described in the previous section, the main effect of the companion in this region is the accumulation effect, which causes the slope of the density in the inner disk to be shallower from the standard VDD steady-state prediction ($m<2$). Thus, compared to a disk of an isolated Be star, we expect a general increase in $V$-band emission and polarization fraction, % (which has a strong dependence on the number of scatterers), 
and in the emission of lines that form close to the star, such as He and Fe lines, \fix{for systems seen at small and intermediate inclinations. These effects should not be visible in edge-on systems}. 

To quantify the effects of this change of $m$ on the spectral energy distribution (SED) of a Be system, one can use the pseudo-photosphere model of \citet{vieira2015a}. The authors find that the spectral slope of the SED in the infrared regime depends strongly on $m$, more so than on the base density of the disk. {When comparing their models to data in \citet{vieira2017}, the authors find that their results could indicate $m < 2$ for several targets that were not confirmed binaries. This is a trend in recent works of spectral fitting of Be stars\footnote{We note that most of the cited works use $n$, the volume density slope, instead of $m$, the surface density slope. They are related as $n = m + \beta$, where $\beta = 1.5$ for isothermal VDD disks.}: \citet{klement2015} and \citet{rubio2023} find $m = 1.63$ and $m = 1.33\pm0.04$, respectively, for $\beta$ CMi, a proposed Be binary\footnote{See \citet{dulaney2017} and \citet{miroshnichenko2023} versus \citet{harmanec2019} for the two opposing views on $\beta$ CMi's binarity}; \citet{klement2019} find the same for their sample of Be stars consisting in known binaries and (assumed) single Be's (see their table 7); \citet{almeida2020} finds $m = 1.5$ for $o$ Aqr; and \citet{marr2021} finds $m = 1.1$ for 66\,Oph.}

%\commentACC{Este parágrafo talvez nas conclusões?}
{The changes in the spectral slope of the SED of Be stars might be signs of binarity, but might also be due to other factors, such as thermal effects not considered in the derivation of the steady-state VDD approximation, radially variable viscosity, and duty cycle of the Be activity \citep{vieira2017}. However, the other regions of a Be binary system also have expected observational effects that could in combination point to binarity more unambiguously.}
This is explored in the following subsections.

% We expect then a general increase in V band emission and polarization fraction (which has a strong dependence on the number of scatterers), and in the emission of lines that form close to the star, such as He and Fe lines.

% \commentACC{Não está correto acima. Certamente as linhas não sao dominadas pelo disco interno. O continuo pode ser, mas apenas para comprimentos de onda mais curtos.}
% We expect then a general increase in V band emission and polarization fraction (which has a strong dependence on the number of scatterers), and in the emission of lines that form close to the star, such as He and Fe lines, but expect no great V/R variation in any lines. Alone, the inner Be disk is a source of inconclusive evidence
% \commentACC{source of inconclusive evidence?}
% where binarity is concerned, since the same effects can be observed in single Be stars with denser disks, or Be stars that are building up their disks \citep{vieira2017}.
% \commentACC{Também não está correto o acima, ver audio.}

\subsection{Spiral dominated disk}\label{sec:spiral}

%A REGIÃO ONDE OS BRAÇOS "COMEÇAM" BATE COM ALGUM RAIO DE RESSONÂNCIA? INCLUIR NOS PLOTS 2D? \commentACC{boa pergunta}
%\ACR{Eu consigo taaaalvez ver alguma coisa em 5:1 pra alpha=1, mas acho q é mt incerto, já q no fim das contas a nossa definição de onde as espirais começam não é exata. do jeito q está, não sei se vale a pena discutir isso}

% \commentACC{a parte do disco interno é desnecessária aqui. }
% As we move further away from the Be star and closer to the companion, its effects on the disk become more and more apparent. Fig. \ref{fig:countour} shows the surface density contours for the entire Be disk (secondary is in the $\phi = 0$ direction). Up until ~XX \req, the density of the disk increases in a azimuthally symmetrical fashion. This region is the inner Be disk (Sect. \ref{sec:inner_be_disk}), where the main effect of the binary is the accumulation of matter, making it denser than the disk of an isolated Be star. The two-armed density waves that are excited in the Be disk by the passage of the companion through periastron, \commentACC{nossas órbitas são circulares, reescrever essa parte para não dar a entender que é a passagem pelo periastro que excita as ondas}
% first predicted in \citet{okazaki2001} (É NESSE MESMO?), begin to become noticeable outwards of XX \req, as seen by the elongation and deformation of the isodensity contours in Fig. \ref{fig:contour}. We refer to this region of the disk, going up to position of the secondary at 18.19 \req, as the spiral dominated disk.
In a quasi-Keplerian disk, as an isolated Be disk should be in theory, the radial velocity should be low, much lower than the sound speed \citep{bjorkman2005}, increasing slowly with radius, but showing no azimuthal dependence. Our simulations of binary Be stars, along with previous works, indicate that in these systems the orbits of the individual particles can be highly perturbed by the secondary's potential. A direct consequence of these changes in the velocity field of the particles is the excitation of two-armed density waves, as seen in Fig.~\ref{fig:velcompare}. %Due to the viscosity and differential rotation of the disk, the

\fix{In this $m=2$ mode, the motion of the particles can be well described by epicycles along their orbit around the Be star, with an epicentric frequency that is twice the orbital frequency, similarly to the motions of stars the disks of galaxies \citep[][Chapter 3]{binney2008}. Thus, there are two apastrons and two periastrons in the orbits of the particles. Thus, the eccentricity measured in Fig.~\ref{fig:grid} (forth row) is instantaneous average of the eccentricity in the wedge.}
The spiral structure, a consequence of this orbital motion, consists of one arm that leads the companion and another, less dense arm that trails 180$\degree$ behind. %Perturbations in the velocity field are a consequence of the orbital eccentricity of the particles (shown in the fourth row of Fig.~\ref{fig:grid}), with 
The consequences for $v_r$ and \vp are shown in the second and third rows of Fig.~\ref{fig:grid}, and are present in all models, with the extreme values for both $v_r$ and \vp happening between the two apastrons and periastrons of the orbit. In the azimuthal direction that includes the stronger density arm (green wedge defined in Fig.~\ref{fig:mapao}), the general behavior of the surface density is of a decrease (much faster than the inner disk), followed by a bump (the arm itself), and finally a more dramatic falloff.
The first decrease has counterparts in the direction of the secondary (yellow wedge) and its opposite direction (blue wedge), and could also be seen in the works of \citet{okazaki2002} and \citet{panoglou2016}. However, given the resolution of their simulations, the second decrease was sharper and more definitive: 
% \commentACC{Corrigir aqui. As simulações do Atsuo+Despina resolviam bem o disco apos os braços, então certamente a primeira queda e o bump são coisas que estão na simulaçõa deles. O que representa algo "blunt e definitivo" seria talvez a ultima queda, após o bump}
it represented a true truncation of the disk. With our updated simulations, we see that what was called the truncation radius is really a region spanning several stellar radii where the density of the disk decreases, but at azimuthally variable rates. Along the green wedge, the density increases again when the stronger density arm (that will become the bridge - see Sect.~\ref{sec:bridge}) crosses it; in the yellow band, the increase happens in the position of the secondary (marked by the dotted vertical lines in the plot); and in the blue band, the increase is much fainter (for some models, such as 30-1.0-0.33 and 30-1.0-0.50, imperceptible in the plot), due to the weaker, trailing spiral arm, expanding away from the system. \fix{In this region, the disk scale height also begins to deviate from the simple power-law expected from a Be disk, as also seen in \citet{cyr2017} in their simulations of misaligned Be binaries.}

\begin{figure}
    \centering
    \includegraphics[width=\linewidth]{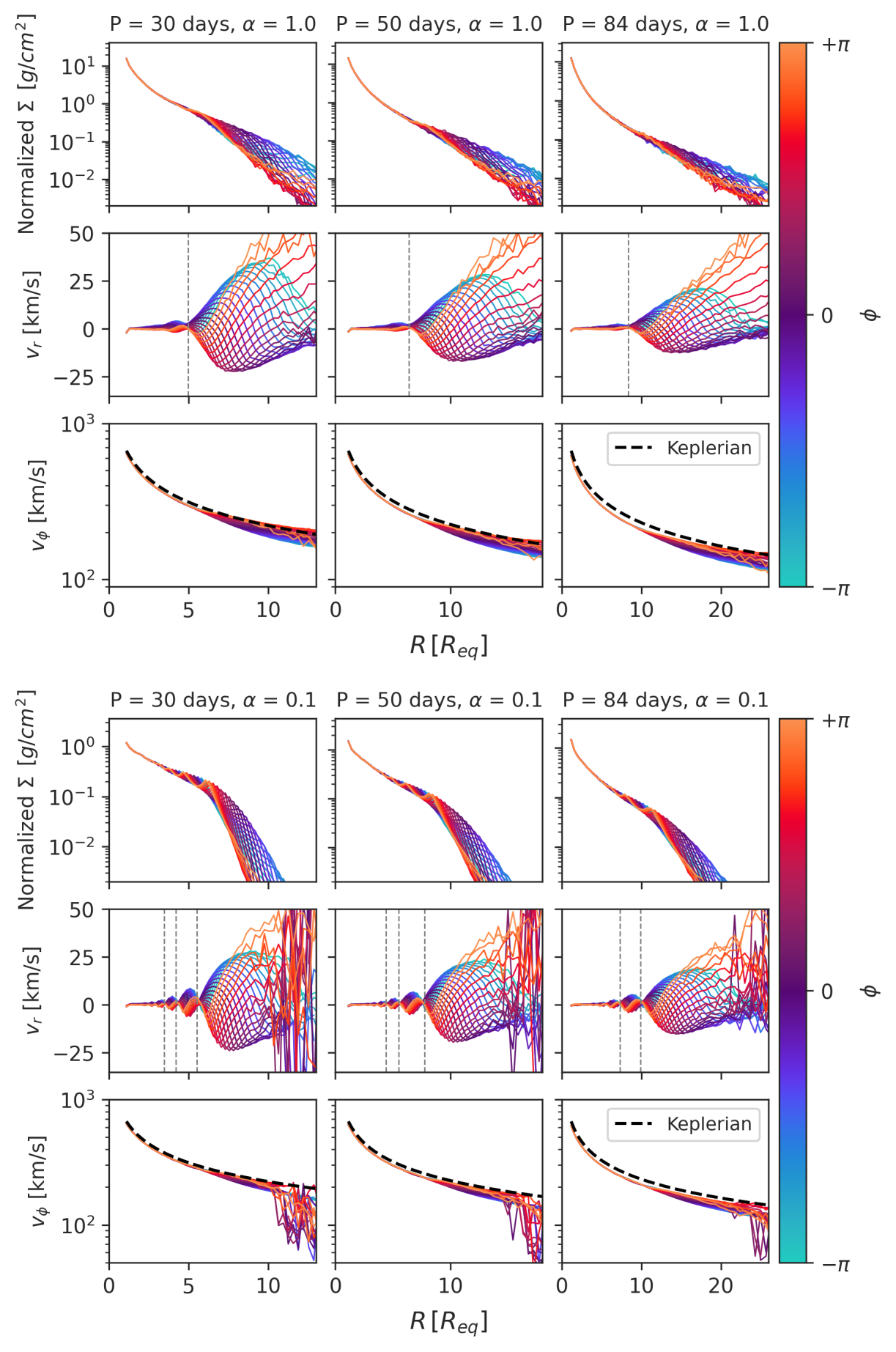}
    \caption{Mean surface density, radial and azimuthal velocities for discrete wedges in $\phi$ of 0.125 radians each (from $\phi$ = -$\pi$ to $\pi$, where $\phi$ = 0 is the direction that includes both the Be star and the secondary). The model parameters are indicated at the top of each column.
    The top and bottom plots compare models with low and high $\alpha$, respectively. Dashed lines in the middle panel represent resonance radii close to the nodes (see text for details). }
    %\commentACC{1) \vp in log? 2) Voce tentou fazer "cunha sim cunha não"? Isso diminuiria o numero de linhas e facilitaria acompanhar linhas individuais. 3) Por consistencia com o texto, não seria o caso de colocar o nome do modelo em cada coluna?}
    
    \label{fig:phis}
\end{figure}

Figure~\ref{fig:phis}, in which each line corresponds to the mean value (averaged over radial bins) of the surface density, $v_r$ and \vp for discrete wedges of 0.125 radians each, offers a more complete view of the dependence of the density and velocity fields on radius and azimuthal angle. The outwardly expanding spiral arms are clearly seen in both density and radial velocity. 

{The effects of binarity on the radial velocity field are especially noticeable, as velocities rise much above the few km\,$\rm s^{-1}$ expected in quasi-Keplerian decretion disks. They surpass the sound speed in the disk and increase even further going towards the companion and into the bridge (Sect.~\ref{sec:bridge}). The reason for this difference is that the radial velocity in isolated Be disks depends only on the viscous transport of matter through the disk. Observations and models agree that Be disks are not wind-driven, but grow through viscous torque and are rotationally supported; therefore, the radial velocity field does not change much with radius for an isolated star. When a companion is added, however, its tidal interaction with the Be disk deviates the orbits of the particles, from the circular and coplanar orbits of an isolated Be disk to elliptical, noncoplanar orbits (see, e.g., the forth panel of Fig.~\ref{fig:grid}). Thus, the significant variations we see in $v_r$ in Figs.~\ref{fig:grid} and \ref{fig:phis} are not due to changes in the rate and velocity of matter ejected from the Be star into the disk, but rather to orbital disturbances caused by the companion. }

% \commentACC{Esse comentário abaixo deve estar lá em cima, na seção anterior (por isso sugeri colocar os valores de $a$ na tabela}
As with the inner disk, the size and behavior of the spiral dominated disk scales with the period of the system. The tightness and ``strength'' of the spirals (the range of their deviation in density -- Eq.~\ref{eq:VDDsigma} -- and velocity from the standard Be disk solution) are, on the other hand, more sensitive to the viscosity. \fix{The emergence of the spiral arms can be seen in the surface density and $v_r$ plots in Fig.~\ref{fig:phis}, particularly for the low $\alpha$ models. The radial velocity exhibits a nodular pattern with points of near-zero velocity aligning with various resonance radii of the orbit. For model 30-0.1-0.16, the resonance radii closer to the visible nodes are 6:1, 9:1, 12:1 and 16:1. For model 50-0.1-0.16, they are 6:1, 10:1, 14:1, and for model 84-0.1-0.16, 7:1 and 11:1. No definite and clear pattern were detected, neither a correlation with the model parameters. %The models follow a 4:1 separation between two consecutive nodes, with the exception of the first two nodes of models 30-0.1-0.16 and 84-0.1-0.16, which have a 2:1 and 5:1 separation (\ACR{NÃO SEI BEM QUAL É O TERMO PRA FALAR SOBRE ESSA SEPARAÇÃO, POR ISSO COLOQUEI 4:1, ETC, Q DEVE ESTAR ERRADO}). 
%\addACC{Sugestão: The nodels all have a }
For our high $\alpha$ models (bottom panels of Fig.~\ref{fig:phis}), only one node is clearly identifiable. The closest resonance radii are 7:1, 8:1 and 9:1 for models 30-1.0-0.16, 50-1.0-0.16 and 84-1.0-0.16, respectively.} %If this trend is real, we expect a binary with a period of 18 days to have this nodule at 6:1, and a binary with 140 days period to have it 10:1.}

\begin{figure}
    \centering
    \includegraphics[width=1.\linewidth]{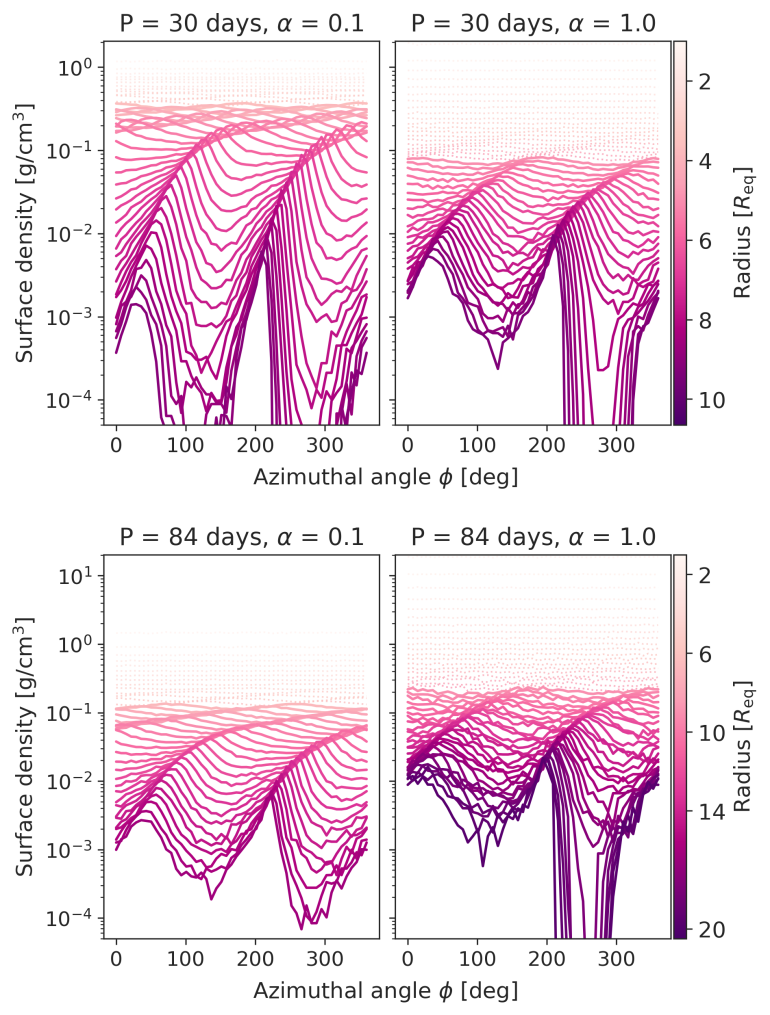}
    \caption{Surface density as a function of $\phi$, at different fixed radial distances from the Be star (shown by the colormap), for models 30-0.1-0.16 and 30-1.0-0.16 (top), and 84-0.1-0.16 and 84-1.0-0.16 (bottom).
    % \commentACC{Melhorar legenda. Aumentar leta do eixo y. Sugiro como antes usar os nomes dos modelos acima das figuras. Pq a escala do modelo abaixo, à direta está somente indo até 1E-3?}
    }
    \label{fig:spirals}
\end{figure}

{Figure~\ref{fig:spirals} offers another view of the spiral arms. Each line represents the surface density as a function of $\phi$, at different fixed radial distances from the Be star (indicated by the color map). 
The tightening and strength of the spirals and the overall size of the region are seen more clearly. Models with larger $\alpha$ have much more spread out and weaker arms, as the density variation is lower. The density variation between the leading and trailing arm is also lower. For the same $\alpha$ and different periods, such as for models 30-0.1-0.16 and 84-0.1-0.16, lower periods show tighter and denser spirals.}

%\commentACC{Seria o caso de mostrar o ajuste da equação abaixo na figura como uma linha pontilhada?} \answerACR{tentei, mas ficou complicado kkk}

To describe the shape of the density waves, we analyzed the surface density curves of Fig.~\ref{fig:spirals}, fitting the azimuthal position of the two peaks with an exponential function
\begin{equation}
    \phi(r) = A \times e^{-\gamma r} + B \,,
\label{eq:winding}
\end{equation}
% \commentACC{verificar a formla acima. A e B estão somando, não Faz sendido pois seriam um parâmetro so. Seria A*exp?}
\noindent where A and B are free parameters {related to the orbital phase and density of the region}, and $\gamma$ is a parameter that measures how tightly wound the spirals are \citep{cyr2020}. The winding parameter $\gamma$ for each model (and for each spiral arm) are presented in Table~\ref{tab:inner}. We find that $\gamma$ consistently decreases with viscosity, as the overall increase of radial velocity in the disk leads to less eccentric particle orbits, and (more weakly) with the period of the system. Increasing mass ratio also leads to tighter spirals. For all models, the spiral arm that leads the secondary is marginally less wound than the trailing arm. This can be also seen examining Fig.~\ref{fig:spirals}. \fix{Our results agree with \citet{cyr2020} for their coplanar models (see their Tables 2 and 3)}.

% \commentACC{DISCUTIR vr! Efeitos de orbita, e nao de outflow. Ver sugestão no final da proxima secao}

\subsubsection{Observational expectations}

The clearest observational counterpart of two armed density waves are the violet/red (V/R) variations in double peaked emission lines, in particular H$\alpha$. These variations are caused by the cyclic increase and decrease of density in either side of the disk as the system rotates and the density waves move from side to side. \fix{While the $m=2$ mode itself is not asymmetric, the two arms are unequal in strength and winding. Furthermore, the optical depth attenuation of the arms depend on the orbit, due to disk flaring; i.e, the arm currently further from the observer will be more affected.}
For Be binaries such as $\pi$\,Aqr \citep{zharikov2013}, $\gamma$\,Cas, 60\,Cyg and HD\,55606 \citep{chojnowski2018}, they phase very closely to the period of the system, but that is not always the case. Recently, several examples of orbital phase-locked V/R variability have been found and measured by \citet{miroshnichenko2023}. Furthermore, they found that the amplitude of maximum V/R value decreases with increasing orbital period (see their Fig.~18). {A quantitative estimate from our models show that there is anti-correlation between the binary separation and the amplitude of the density variation in the spiral dominated disk, which likely translates observationaly as the V/R amplitude-period relation detected by \citet{miroshnichenko2023}}.

One-armed density waves are also excited in Be disks with no relation to their binarity \citep{okazaki1991}, {leading to V/R variations with much larger amplitudes and generally much longer periods \citep[typically longer than a few years as the period is set by the precession the of the one-armed wave around the star, e.g., $\zeta$\,Tau,][]{carciofi2009}. 
The $m=1$ waves, therefore, give rise to phenomena similar to those of the $m=2$ waves, but can be easily differentiated by their longer periods and larger amplitudes.

How tight and how strong the spiral arms are will influence its effects on the observables, which also depend on the inclination angle of the system. Larger V/R variations will happen when there is stronger left-right asymmetry in the disk with respect to the observer. 
{More spread out waves will produce larger V/R variations, while this asymmetry will clearly be smaller (washed-out) for more tightly wound spirals, as they decrease the density and velocity contrast in the emitting area of the disk.

\citet{panoglou2018} studied the expected behavior of H$\alpha$ line profiles by calculating radiative transfer simulations for systems of coplanar, circular Be binaries, based on the SPH simulations presented in \citet{panoglou2016}. Their models agree with our expectations for the general behavior of the V/R variations caused by the density waves, but we note that their radiative transfer calculations are based only on the density structure of the perturbed Be disk given by their SPH simulations: their models do not consider the velocity field of the particles, and include only what we refer to as the inner Be disk and the spiral dominated disk, fully truncated by the companion. Thus, any observational effects of the bridge, circumsecondary region, and circumbinary outflow, and the relevant effects of the velocity profiles of the particles, are not considered in their H$\alpha$ calculations.

{Effects in the polarization are also expected to arise from the disk perturbations. Variation in the position angle (PA) of $\zeta$ Tau with phase due to $m=1$ density waves is well established \citep{schaefer2010}. Similar effects have been predicted to exist in the case of $m=2$ waves of Be binaries by \citep{panoglou2019}. However, the observational detection of this phenomenon remains to be achieved.

An effect that should also be seen in binary Be stars is variations in the position of the central reversal in double peaked lines, in particular for shell lines. Shell line profiles are characterized by unusually deep central reversals, caused by absorption of stellar light by the disk, and appear in Be stars observed at high inclination angles (closer to edge-on). 
In symmetric disks, with small radial velocities %close to 0 km $s^{-1}$ 
%are responsible for this deep central reversal. The 
the shell profile is symmetric, narrow and centered at the systemic velocity of the star. For disks with asymmetries such as the ones % caused by density waves (both one and two-armed) due to the previously discussed orbital effects, 
discussed here, 
the projected velocity along the line of sight can be rather large, as shown in \fix{Fig.~\ref{fig:projvels} for model 84-0.5-0.16 and observes at two different positions ($\phi = $18\degree -- right, and 350\degree -- left), indicated as the dashed lines}.
%. Figure~\ref{fig:projvels} shows the projected velocities along the light of sight of the observerui of our model 84-0.5-0.16 for an observer at $\phi = $18\degree \,(right) and 350\degree \,(left). 
The regions corresponding to different projected velocities %(positive and negative) 
are marked in different colors.
\fix{The light emitted by the star travels through regions with quite different radial velocities before reaching the observer, in stark contrast to a symmetric disk model. Therefore, it is expected that the shell absorption will not necessary be centered at the systemic velocity, and will likely be broader due to the larger range of radial velocities.}
\fix{This effect was observed in the Be+sdO binary HD\,55606, which goes through orbital phase dependent intermittent shell phases.} 
%\commentACC{Este precisa ser reescrito. A frase We attribute é uma repeticão desnecessária. Talvez algo assim para começar}
%\addACC{The above effects on the central reversal may be present on non-shell stars as well.}
% \fix{The effects on the central reversal due to changes in the velocity field of the disk are not limited to shell stars.}
\citet{chojnowski2018} shows that the wings of the H$\beta$, H$\gamma$ and H$\epsilon$ lines of this system do not shift with phase, while the shell feature does (see their Fig. 7). %We attribute this central reversal variation to the asymmetric projected velocities of the disk with respect to phase, due to the asymmetries in the disk caused by the companion, \fix{as shown in Fig.~\ref{fig:projvels}}. 
This central reversal variation has been little explored in the literature and offers great diagnostic potential in characterizing the properties of binary Be stars.

\subsection{Bridge}\label{sec:bridge}

\begin{figure*}
    \centering
    \includegraphics[width=\textwidth]{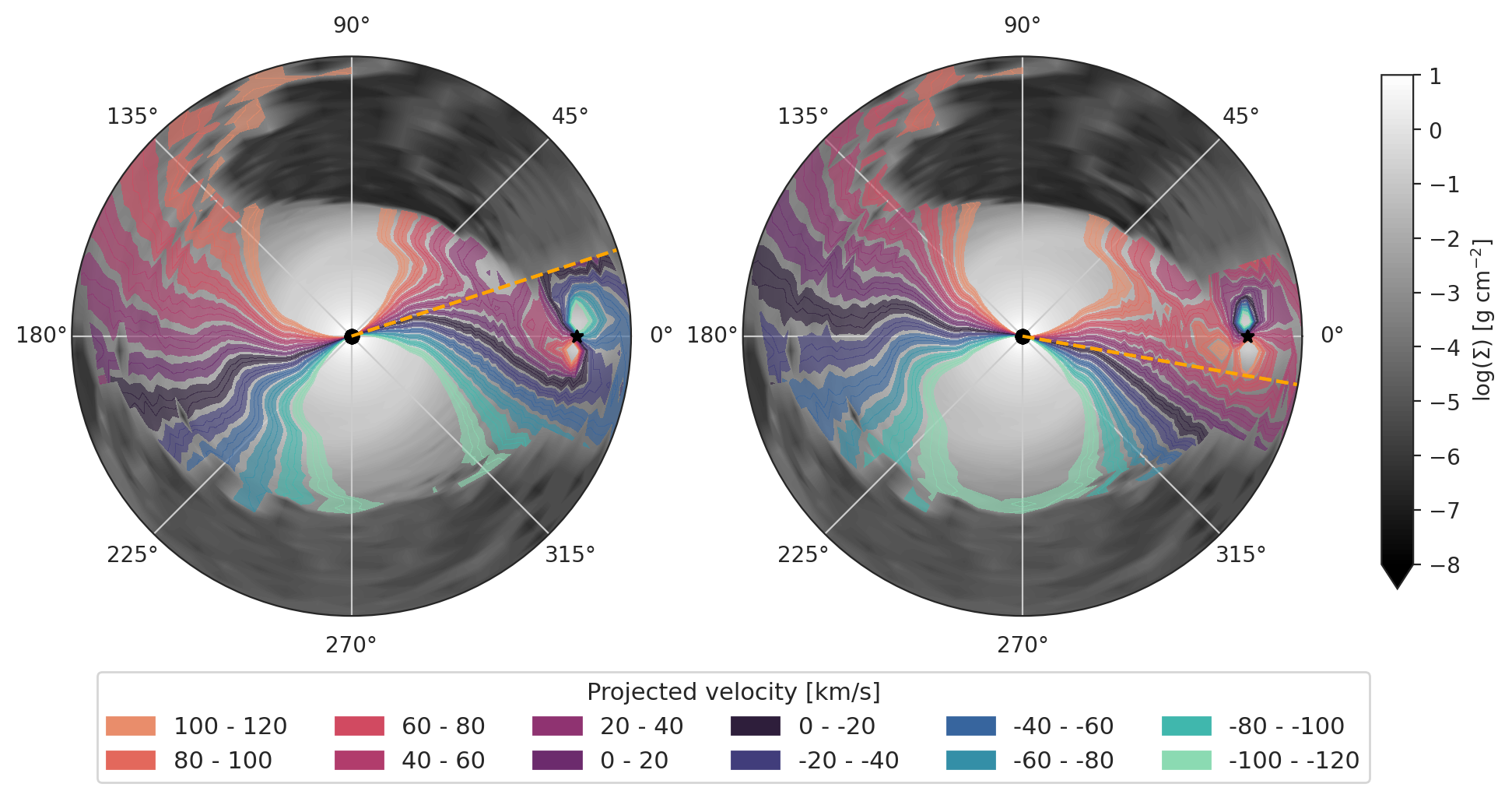}
    \caption{Projected velocities along the line of sight of an observer at $\phi=$ 18\degree (left) and $\phi=$ 350\degree (right), for model 84-0.5-0.16. The background black and white map show the surface density of the system. The various colored bands map out the projected velocities, while the dashed orange lines indicate the direction of the observer. \revi{The black circle denotes the Be star, and the black star, the companion.} }
    \label{fig:projvels}
\end{figure*}

% \commentACC{Sugiro uma mudança de estrutura. Nesta seção, discutir apenas a ponte de forma qualitativa -- usar textos da seção anterior que ficaram melhor aqui. Para a parte de BHL e RLOF, colocar uma seção para isso ao final - secao 3.5 - Accretion onto the Secondary}

Along with the accumulation effect and the spiral arms, the truncation of the Be disk is a known effect of binarity. As the disk grows, the resonant torque of the companion takes its angular momentum, effectively impeding its further growth. In the SPH simulations of \citet{okazaki2002} and \citet{panoglou2016}, this truncation happens at around the 4:1 to 5:1 resonance radii and is rather sharp, given their larger size of the secondary's sink (already discussed in Sect.~\ref{sec:prelim}). Our updated simulations show a different perspective: the disk is not totally truncated, but is elongated along the Roche potentials of the binary system.
{The disk scale height in this truncation region, shown in the bottom panels of Fig.~\ref{fig:grid}, behaves very differently depending on the azimuthal angle: the decrease is sharp for the $\phi = 0$ wedge (that contains the secondary) but is not present in the opposite direction ($\phi = 180$\degree wedge), {except for a subtle decrease in models 30-1.0-0.33 and 30-1.0-0.50. Note, however, that, given the low resolution of the system towards $\phi = 180$\degree,  it is difficult to fully characterize the scale height in this direction.} 
% \commentACC{Amanda, precisamos reescrever acima, não há decrescimo de H na cunha de 180...}

Our models show that matter flows from the Be disk via two exit points, one towards the companion and another opposite to it. 
These exit points are located around the L1 and L3 Lagrangian points (see Fig.~\ref{fig:densmaps}), but, given the rotation of the system, they are leading L1 and L3 in the direction of rotation.
% \commentACC{reescrever acima indicando nuance importante: o exit point é "leading" com respeito aos L's}
The surface density maps of Fig.~\ref{fig:densmaps} for models 30-1.0-0.16, 30-1.0-0.33 and 30-1.0-0.50 show that the mass ratio of the system not only affects the overall size of the Be disk, but also the efficiency of mass leakage through L3. The same effect can be seen in the bottom right panel of Fig.~\ref{fig:grid}, where the scale height of the disk decreases more or less equally in all directions for models 30-1.0-0.33 and 30-1.0-0.50, suggesting a more evenly truncated disk than seen for model 30-1.0-0.16.}

\begin{figure}
    \centering
    \includegraphics[width=.85\linewidth]{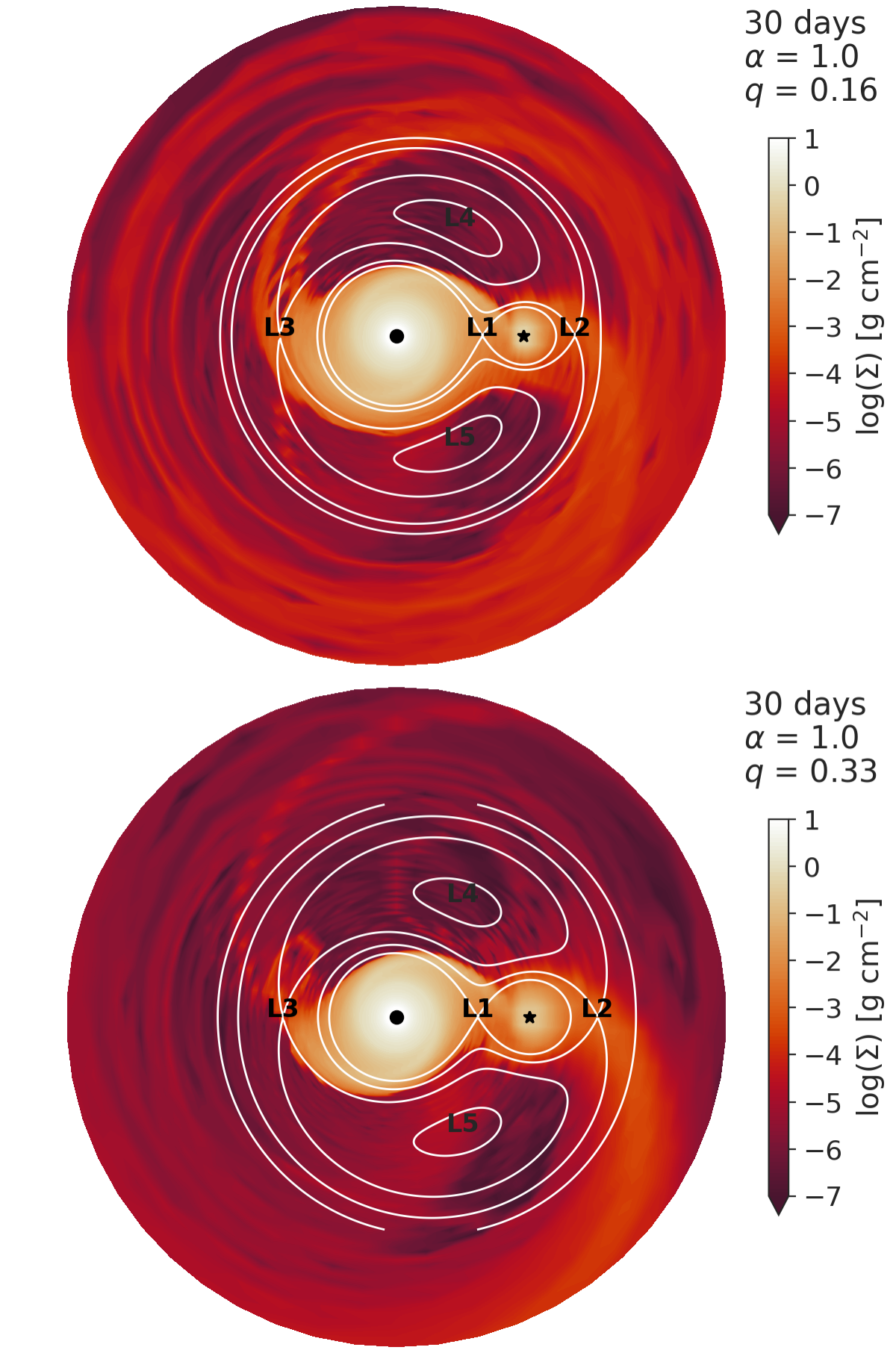}
    \caption{Density maps for models 30-1.0-0.16 and 30-1.0-0.33, showing the Roche equipotential contours and the Lagrangian points. \revi{The black 
    circle marks the Be star, and the black star, the companion.}
    %\commentACC{Usar os nomes dos modelos na Fig. Talvez tirar a 3a? Ela é muito parecida com a segunda.}
    }
    \label{fig:densmaps}
\end{figure}

The transfer of matter from the Be disk into the Roche Lobe of the companion happens through the bridge, an extension of the leading spiral arm that connects the two (light green region in Fig. \ref{fig:esquema}). Most of the mass that leaves the disk of the Be star does so through the bridge, %, formed by a combination of the tidal torque of the companion and the rotation of the disk.
which is by far the densest region at this radius (for instance, in model 30-1.0-0.16 the bridge is seen as the density peak at about 12 \req, green line in the top right panel of Fig.~\ref{fig:grid}). It is also the region with higher values of radial velocity compared to the rest of the Be disk, as the material is accelerated towards the companion and then enters its RL close to the L1 point. %\commentACC{mesmo cuidado acima, não é pelo L1}. 
Its' counterpart, the trailing arm opposite to the companion, is the second point of mass loss for the Be disk, near L3. This arm grows outwards undisturbed, eventually merging with the stream that forms the circumbinary region (Sect. \ref{sec:sec_struc}).%, but is absent is some models (Fig.~\ref{fig:densmaps}).

\begin{figure}
    \centering
    \includegraphics[width=1.\linewidth]{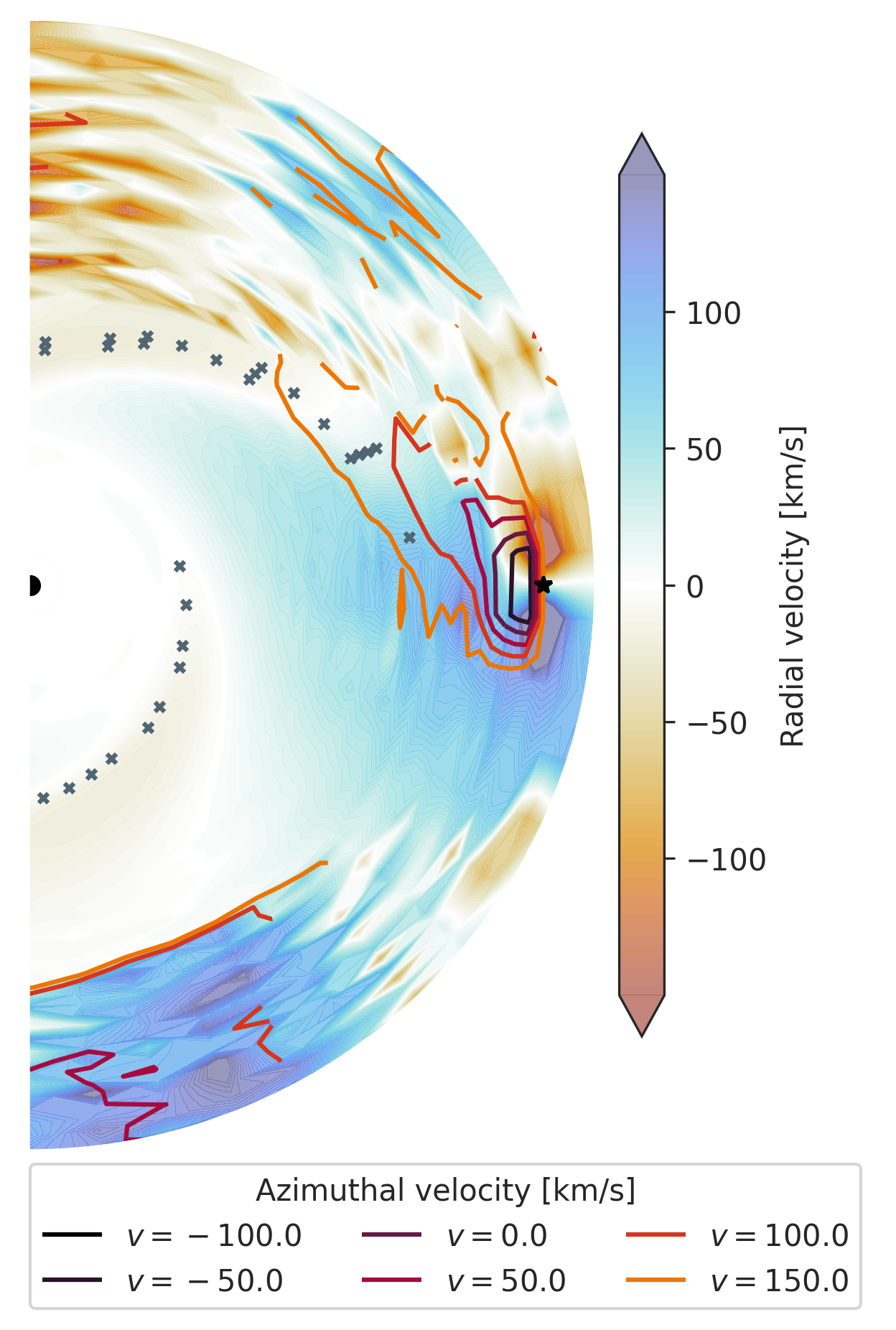}
    \caption{Map of radial (in full colors) and azimuthal (contours) velocity for simulation 30-1.0-0.16. The gray x's mark the regions of highest density for that given radii, tracking the two spiral arms. The black half circle marks the Be star, and the black star denotes the companion.}
    \label{fig:cont}
\end{figure}

% \rmACC{The bridge is characterised by enhanced density (green lines in the top panels of Fig. \ref{fig:grid}) and significant changes in the velocity field. The radial point of highest density in the bridge changes slightly with mass ratio, and for models with the same parameters, the density decreases with $\alpha$.} 
Figure~\ref{fig:cont} shows a zoomed-in 2D map of the bridge for model 30-1.0-0.16, with the color indicating the radial velocity of the gas (in the reference frame of the primary). 
Since the radial velocity of the secondary is zero (due to its circular orbit), the plot illustrates the significant velocity differences between the secondary and the surrounding gas as material accelerates along the bridge towards the companion. The positions of the two density arms are marked by the x's in the plot. Superimposed on the image are contour lines indicating the azimuthal velocity. These lines reveal that the Be disk is now far from Keplerian, even exhibiting negative azimuthal velocities, indicating retrograde motion.
%The high azimuthal velocity in the Be disk decrease and even go to negative values in the bridge, and the radial velocity, low in the spiral arm that originates the bridge (marked by the gray xs), increases as material accelerated towards the companion. 

\subsubsection{Observational expectations }\label{subsec:bridgeobs}

Observationally, the steep density fall-off of the truncation region is seen as a steepening of the slope of the spectral energy distribution (SED) in long wavelengths, around the centimetre/millimetre regions. This effect is referred to as the SED turndown \citep{waters1991, klement2017}. % and is \rmACC{a consequence of the decrease of the emitting area of the disk.}\commentACC{cuidado aqui, o motivo nao é decrescimo}
SED turndowns have been detected in several Be stars in \citet{klement2019}, both in confirmed binaries (e.g., $\gamma$ Cas) and in (at the time) presumed isolated stars (e.g., $\beta$ CMi). As previously discussed, confirming binarity in Be stars is a complex task due to the brightness ratio between the Be and its usually evolved companion, and to the effects of stellar rotation and of the disk on their spectral lines. The detection of the SED turndown, while an indicator of disk truncation, does not automatically mean a companion is present, as other causes for truncation have not been discarded (see, for instance, the results of \citealt{cure2022} for the transonic disk solution, and \citealt{kee2016} for radiative ablation of the disk). 
% \rmACC{The overall size of the disk can also be estimated, roughly, from the equivalent width of the H$\alpha$ lines \citep{grundstrom2006}.} 
Our models show that in reality it is more complex: the extent of the truncation region varies with azimuth due to the disk's elongation. Additionally, the bridge acts as a funnel through which most of the disk mass flows, complicating matters by extending a small part of the disk to larger radii.

{\citet{peters2016} provides a prime example to understand the impact of the bridge on the observations.} They assumed a ``circumbinary disk'' model to explain the shell absorption features in \ion{Si}{II} and \ion{}{III} and \ion{S}{III} lines of HR2142, a known Be+sdO binary. Their model is close to what is shown by our simulations: a disk truncated by the companion, but where material travels through the L1 point towards the companion in streams (similar to what we call the bridge in this work). The opacity of the stream would be responsible for the shell absorption as the stream passes in front of the Be star. \citeauthor{peters2016} also predicted the leak of material through L3, and the formation of a circumbinary disk, in agreement with our models.

In \citet{peters2016}'s observations of HR2142, the shell features occur in two phases. {In the what they called the primary phase (when the observer is at around $\phi$ = 18\degree), the absorption features span a velocity range of +10 to +160 km $s^{-1}$; in the secondary phase (observer around $\phi$ = 349\degree), velocity ranges from -40 to +20 km $s^{-1}$.} To investigate whether these phases are compatible with what we expect from the bridge in our simulations, we show in Fig.~\ref{fig:projvels} the projected velocities for observers at $\phi = 350.0$ and $\phi = 18.0$ (with the bridge in front of the Be star) for model 84-0.5-0.16. As the figures show, there is indeed material in the bridge and circumsecondary region (see Sect.\ref{sec:sec_struc}) with radial velocities roughly matching the ranges seen by \citet{peters2016}. Other disturbances on line profiles due to the bridge can also appear. Whether this contribution is in emission or absorption depends on the inclination angle of the observer: absorption features are expected for near edge-on orientation and vice-versa. Here the nature of the companion becomes important, as this region is close enough to be irradiated by it. If the companion is a hot subdwarf, as is the case for many Be stars, including HR2142, it is possible that the bridge
will be further heated up by the secondary, changing the emission/absorption line strengths.
Traveling features (in absorption or emission) can, therefore, be caused by the bridge, and might be indicators of binarity. 

%\commentACC{cuidado com as explicações acima usando o estado de ion., etc. Se quiser depois conversamos sobre isso, a coisa é mais complicada}

% \ACR{We'll add a figure here with predictions for the projected velocities of the bridge}

\subsection{Circumsecondary region}\label{sec:sec_struc}

 % Material that comes from the bridge with high radial velocities is decelerated and begins to orbit, or at least bend, around the companion, leading to an increase in density, and, finally, to accretion.

Via the bridge, material from the Be disk enters the RL of the secondary. Given the boundary conditions previously set in the SPH code of \citet{okazaki2002}, no structure of any kind could be seen around the secondary in their Be binary simulations. % due to low resolution in the outer Be disk. 
\citet{hayasaki2004} bypassed this issue by running a simulation focused only on the accretion stream (the bridge). The boundary conditions for this simulation were chosen based on the parameters of the particles that were accreted by the companion (i.e., entered its RL) in one of the simulations of \citet{okazaki2002} (based on a coplanar BeXRB system with $e = 0.34$). Their results show that a Keplerian disk is formed around the companion for all their simulations, which explored different equations of state and viscosity parameters. \citet{martin2014}, using SPH simulations with the code \textsc{phantom} \citep{price2018}, also shows that an accretion disk is formed. These works investigated tilted and eccentric systems for which there is an extended period of time where no new matter and AM enters the RL of the companion. During this time, the matter already inside the RL, dumped there in the previous encounter with the Be disk, can reorganize itself and form a disk from which the companion accretes. We note that these works keep the artificial SPH viscosity ($\alpha_{\rm SPH}$) fixed and do not considered the effect of the secondary on the scale height of the disk, which lead to increasing Shakura-Sunyaev $\alpha$ in the vicinity of the companion (with the exception of \citet{hayasaki2004}: as it considered only the region of the gravitational sphere of the secondary, where the particle distances are measured from the secondary, so no artificial increase of $\alpha$ occurred). Our simulations use a prescription for the viscosity of Be disks in binary systems where the SS $\alpha$ does not vary in time or with radius (see Sect.~\ref{sec:viscpresc}).

In our simulations, from the moment the Be disk reaches a quasi steady-state configuration (see Sect.~\ref{sec:temporal_evol}), there is a constant injection of mass and AM into the RL of the companion. The velocity and density of this particle stream, as discussed in Sect.~\ref{sec:bridge}, are parameter dependent, and so is the circumsecondary region. The yellow lines in Fig.~\ref{fig:grid} show the properties of the material in the yellow wedge of Fig.~\ref{fig:mapao}, a region that contains the secondary and the majority of its RL. There is a tremendous increase in density in this area, even rivaling the mean density of the spiral dominated regions of the Be disk. The same is also shown in the cross section contour plots of the volume density (considering a height $z = -10$ to 10 \req) of the circumsecondary region for all models are in Fig.~\ref{fig:densmaps_cont} (black lines). The orbital period has the strongest effect on the density of the circumsecondary region: lower periods lead to denser structures.
The effect of $\alpha$ is more clearly seen in the velocity profile, where lower viscosity models have a larger rotational velocity, exceeding 20 times the radial velocity, in the whole circumsecondary region. This is likely due to the fact that in large $\alpha$ models the material comes from the bridge with higher radial velocities and has higher viscous torque.
% However, $\alpha$ also has a significant effect, as lower values produce circular structures, while higher viscosity leads to a more extended structure. This is likely due to the fact that material comes from the bridge with higher radial velocities \fix{and has higher viscous torque}, and is thus less easily captured. 
%\commentACC{Não acho que a explicação acima esteja certa, pois neste caso não se formaria uma estrutura com simetria circular}
For higher mass ratios (bottom rows of Fig.~\ref{fig:grid}), the concentration of matter is larger, and so is its rotational velocity, as a more massive companion has a stronger gravitational pull.
%\commentACC{Amanda, um problema que aparece imediatamente na figura é o valor maximo da densidade, 1.E-13, o que da no máximo 1.E11 em denisdade de particula. Isto não gera nenhum efeito observável, pois no HDUST todos os modelos que rodei com n < 1.E12 não produziam nenhum sinal polarimetrico ou espectroscopico. O que me ocorre é ver como está a sua normalização da densidade. Qual a densidade de base da estrela Be? Talvez isso pudesse ser aumentado.}

%However, for higher $\alpha$ and $q=0.16$, the radial velocities of the particles are of the order of hundreds of km $s^{-1}$, and since there is no relaxation time, the orbits of the particles cannot circularise, and the disk is deformed. For models 30-1.0-0.33 and 30-1.0-0.50, however, the higher mass of the companion captures particles more efficiently, forming disks with low radial velocities that are closer to Keplerian. 

% \ACR{Show maps of the velocities in an zxy=0 cut?}

%\ACR{Dar uma definição mais quantitativa do que estamos considerando um disco.}
% \commentACC{Podemos considerar disco tudo que a velocidade orbital seja maior que a velocidade de inflow ou outflow}

\begin{figure*}
    \centering
    \includegraphics[width=0.7\linewidth]{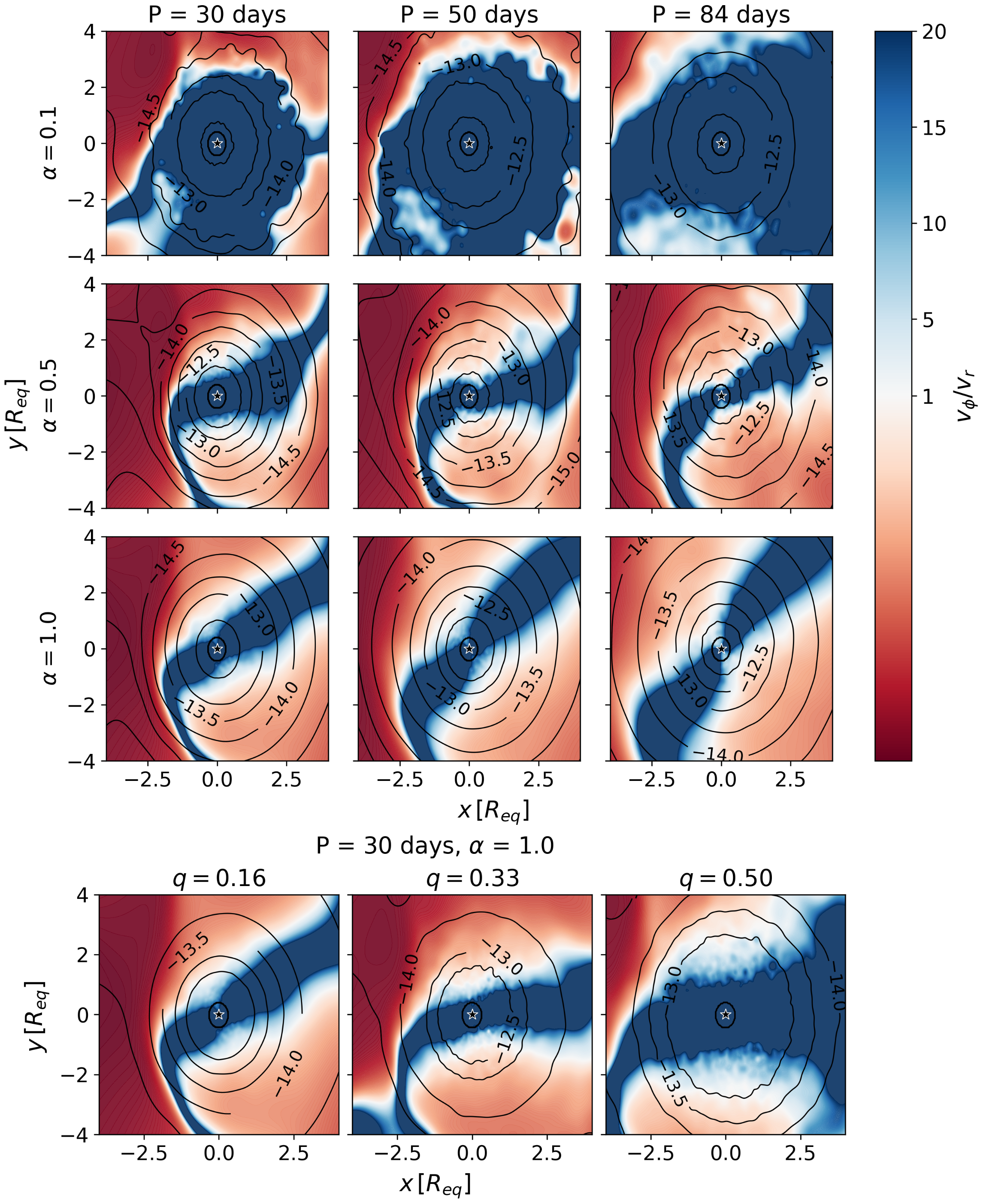}
    \caption{Map of the ratio of $v_{\phi}$ to $v_r$ in the reference frame of the secondary, indicated by a black star in the center. Cross section contour plots (at $z = 0$) of the volume density (log($\rho$), in g cm$^{-3}$, integrated from $z = -10$ to 10 \req) of the circumsecondary region are shown in black. The Be star is to the left, not shown.}
    \label{fig:densmaps_cont}
\end{figure*}
 
% \commentACC{Não é isso que a nova Fig. 11 mostra. TModelos de alpha grande mostra que a região que é dominada pela rotação é pequena. Então é bom rever o parágrafo abaixo. Eu conduziria a discussão falando que um disco está present quando a região dominada pela rotação (azul) circunda totalmente a secundária. Isto é verdade paranas para 4 modelos na figura}
% {In all of our simulations, the structure has a quasi-Keplerian rotational profile (in the reference frame of the companion). 

We consider the circumsecondary structure (where particles are gravitationally bound to the secondary) to be disk-like (i.e., dominated by rotation -- blue region in Fig.~\ref{fig:densmaps_cont}) when its orbital velocity is larger than its inflow and outflow radial velocities, in the frame of reference of the secondary. Fig.~\ref{fig:densmaps_cont} shows the ratio of the azimuthal and radial velocity fields for all models. The deformation of the structure in the higher $\alpha$ models, and their densities are lower when compared to models with $\alpha=0.1$. For models with higher mass ratios, the rotational component of the velocity is very significant, even more so than for low viscosity models. Even so, all models show a large portion of their circumsecondary regions to be dominated by rotational velocity, and show circular symmetry in the density. We conclude that the structure can be considered disk-like in all our simulations, keeping in mind that its size and density are highly dependent on the system's orbit. 
\revi{We note again that our simulations are purely hydrodynamical and isothermal, lacking radiative transfer.} Consequently, the actual structure of the circumsecondary region, particularly very close to the companion, cannot be fully determined at this time. Additionally, it may be important to consider the winds of stripped stars in this context.

% \begin{figure*}
%     \centering
%     \includegraphics[width=1.\linewidth]{figs/splasharrows.png}
%     \caption{Cross section volume density maps for simulations 30-1.0-0.16 (left) and 30-0.1-0.16 (right). Arrows indicate the velocity of the material relative to the companion, marked by the black circle. The Be star is to the left (not shown), and part of the bridge can be seen in the upper left corners of the two maps.
%     \commentACC{1) LEtras pequenas demais. 2) Acima voce sempre usa menor alpha à esquerda e maior a direta (Ex. Fig 7). Sugiro manter semrpe a mesma ordem, ajuda o leitor.}
%     }
%     \label{fig:arrows}
% \end{figure*}

% The growth of the circumsecondary structure is dependent on the viscosity parameter. Once the disk is big enough so that the spiral arms and the bridge are formed and matter is constantly being accelerated towards the companion, the role of the viscosity is still significant in determining the shape and behaviour of the structure that forms around the companion and the mode of its accretion: is the structure around the companion more similar to an outflow or an accretion disk, and if it is a disk, what is its velocity structure?  

Particles are continuously accreted by the companion throughout the simulation, as shown in Fig. \ref{fig:acrates}. The dependence of the accretion rate by the companion ($\dot{M}_{\rm acc}$) with the evolution of the simulation is due to two main factors: the time necessary for the disk to grow and reach the orbit of the companion, and the size of the structure formed around the companion as matter accumulates inside its RL. The amount of days of constant mass injection needed for $\dot{M}_{\rm acc}$ to grow above $1.3 \times 10^{-11}$ $M_\odot$/year (an arbitrary definition based on the inflection points of $\dot{M}_{\rm acc}$ in Fig. \ref{fig:acrates}) scales with period, $\alpha$, and mass ratio, and can be viewed as a rough estimate of how many binary orbits it takes for the Be disk to build-up completely in each model. These estimates are given in Table~\ref{tab:inner}. As expected, the Be disk build-up time decreases with increasing $\alpha$ (Eq. \ref{eq:visctimescale}), but are not consistent across different periods and mass ratios, which might indicate it is due to SPH uncertainty rather than being a physical property. \revi{We also emphasize that these estimates are for the growth of Be disks subject to constant mass injection from the Be star, and depend on the mass injection rate, $\dot{M}_{ij}$. As discussed in Section \ref{sec:models}, $\dot{M}_{ij}$ in our simulations is fixed to $10^{-8} M_\odot \rm{yr}^{-1}$, which also fixes the particle mass in each simulation.  }

% \ACR{Re-checking this point}
% \commentACC{Hmmmm.......} \answerACR{probably an indication?}

The most commonly used physical descriptions of mass accretion are the Bondi-Hoyle-Lyttleton (BHL) \citep[see][for a review]{edgar2004} flow accretion and Roche lobe overflow (RLOF). BHL considers accretion onto a star from a supersonic, infinite, homogeneous gas cloud, and is mostly used for accretion of winds. RLOF is used in the context of binaries, where one star inflates and fills its Roche lobe, initiating mass transfer. In RLOF, mass flows through the Lagrangian point L1 onto the RL of the companion, and is then accreted.The case of Be binaries exhibits clear similarities to both BHL and RLOF accretion scenarios. As in BHL, matter from the Be disk flows towards the secondary with high radial velocities; and as in RLOF, the disk expands enough to reach the RL limits. As seen above, mass transfer does not happen exactly through the L1 point as a regular RLOF, but ahead of it, due to the non-zero rotational component of the disk at the disk/companion interface, and due to tidal friction between the Be disk and the tide-raising secondary. 

%tidal friction is the reason why the tidal bulge moves ahead of the tide-raising secondary. If there is no viscosity, the tidal bulge responds instantaneously to the tidal torques. But, there is viscosity, so the response delays, Since particles in the tidal bulge (Be disk) rotate faster than does the secondary, the tidal deformation occurs slightly after they pass the closest positions to the secondary. 

The BHL accretion rate is given by
\begin{equation}
    \dot{M}_{\rm BHL} = \rho v_{\rm rel} \pi b^{2},
\end{equation}
\noindent
where the impact parameter $b$ is defined as $b = {2 G M_2}/{v_{\rm rel}^2}$, and $\rho$ and $v_{\rm rel}$ are the relative density and velocity of the flow. In the case of winds, where the BHL prescription is commonly used, these quantities are defined in a region where they are not yet affected by the gravitational pull of the accretor. However, these quantities are not as straightforward to define in the case of a accretion from a Be disk. Even in the case of an isolated Be star, the decretion disk has radially and vertically dependent density and velocity profiles. For a disk perturbed by a companion, these profiles become even more complex %dueto the viscous forces on the system 
(as described in previous sections). Therefore, there is virtually no location in the disk where the presence of the companion is not affecting the flow of the material.

% The BHL prescription does not account for the viscous forces acting on the system, and the non-uniform structure of the disk make it difficult to apply it directly to the case at hand. However, we can estimate the what the BHL accretion rate would be if we consider the density and velocity structures of the material on the bridge. As shown in Fig. MAPS, the mean volume density in the brigde is $10^{-15} \mathrm{g/cm^3}$, and the relative velocity to the companion is on the order of 100 km $s^{-1}$. Considering also that only the material inside the RL can be accreted, even if the impact parameter $b$ is larger,  $\dot{M}_{\rm BHL}$ between $10^{-9}$ and  $10^{-10} \, M_{\odot}/\rm{year}$, of the same order of magnitude we find in our SPH simulations, close to $10^{-9} \, M_{\odot}/\rm{year}$ (Fig. \ref{fig:acrates}). 

The BHL prescription does not account for the viscous forces acting on the system, and the non-uniform structure of the disk make it difficult to apply it directly to the case at hand. However, we can (crudely) estimate what the BHL accretion rate would be considering the density and velocity structures of an isolated Be disk, if a companion instantaneously appeared in the middle of the disk at a distance from the Be star at a given separation, with no time given for it to alter its density and velocity structures. Assuming a Keplerian disk, the relative velocity $v_{\rm rel}$ between the disk material and the companion in orbit around the Be star will be the radial velocity of the disk at that radius, which are of the order of tens of $\rm km\,s^{-1}$. The density will be given by Eq. \ref{eq:VDDrho}. Considering also that only the material inside the RL can be accreted, even if the impact parameter $b$ is larger (which is the case, given the low relative velocity), we find BHL mass accretion rates of the order of $10^{-11} M_{\odot}/\rm{year}$, about 20 times smaller than we find in our SPH simulations ($\sim 2 \times 10^{-10} \, M_{\odot}/\rm{year}$).

The overall shape of the accretion stream in our coplanar Be binaries is nearly identical to wind-RLOF \citep[WRLOF -- see][]{mohamed2007}, where wind, and not the star itself, is what fills the RL. In WRLOF, accretion rates can exceed wind BHL prediction by $\approx$100 times, against the 20 times larger estimate we find in our ``disk-RLOF''. This discrepancy is caused by the viscous forces acting on the system and the disk rotation. The bridge has an azimuthal velocity lower than Keplerian, as the gravity well of the companion decelerates the material azimuthally, while accelerating it radially. This leads to a larger relative velocity between the bridge material and the companion, making capture and accretion more difficult.

%The dependency of the accretion rate of the secondary $\dot{M}_{\rm acc}$ with the viscosity parameter $\alpha$ is due to the iron grasp that $\alpha$ has on the growth rate of the Be disk. The lower $\alpha = 0.1$ creates a disk that is slow to build up, while $\alpha = 1.0$ builds it 10 times faster. 

\begin{figure}
    \centering
    \includegraphics[width=0.95\linewidth]{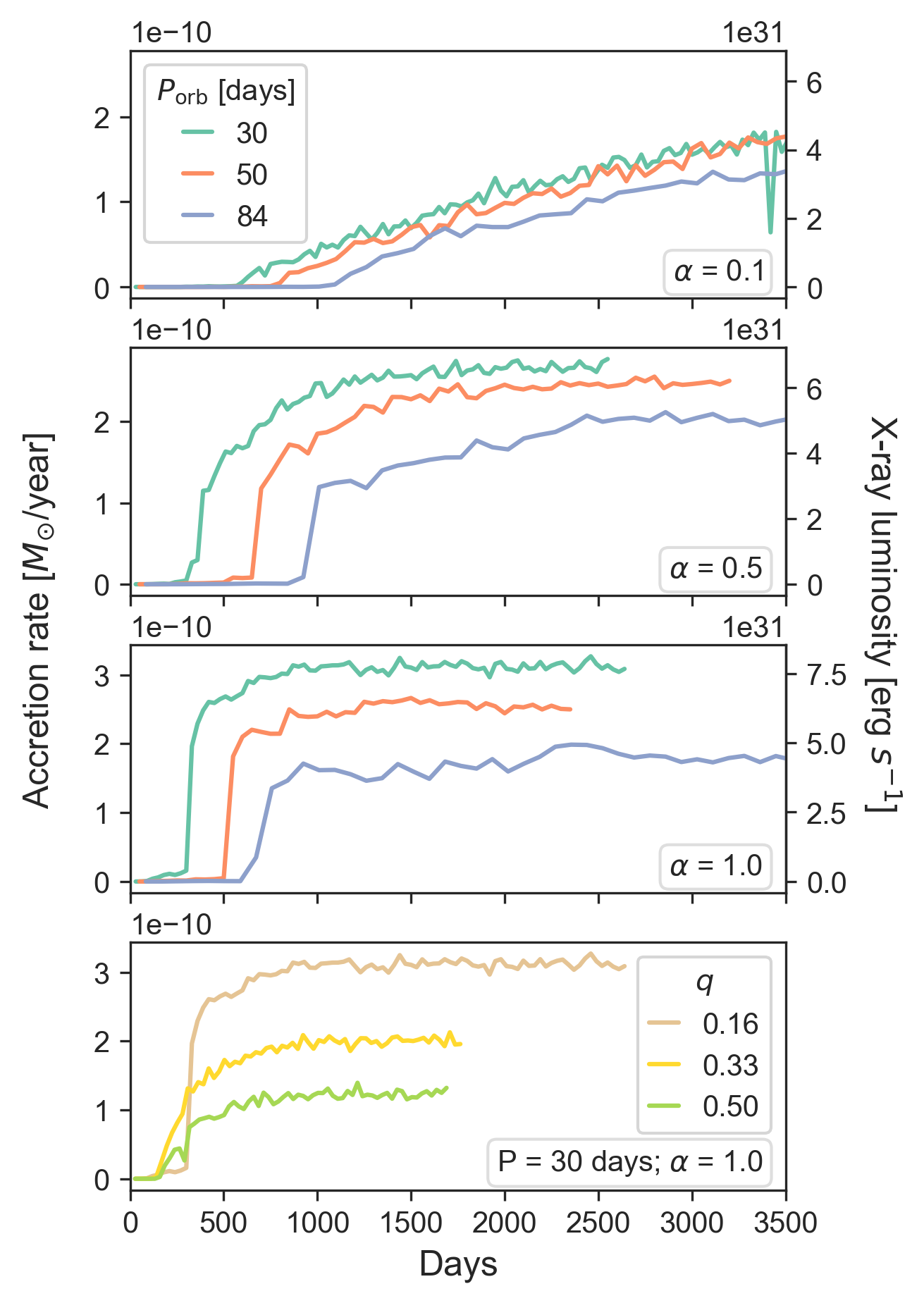}
    \caption{Accretion rates and X-ray luminosities for our models. In the first three panels, different colours indicate period, and $q$ and $\alpha$ remain fixed. In the forth panel, colours indicate $q$, and period and $\alpha$ are fixed.}
    \label{fig:acrates}
\end{figure}

\fix{The maximum value of $\dot{M}_{\rm acc}$ strongly depends on the orbital period and on $\alpha$. In our models, the Be disk is fed constantly, and the rate of mass flow eventually reaches a steady-state.% when the disk grows enough. 
This steady-state indicates that the amount of matter injected and ejected from the system (leaving either via accretion onto the secondary or to the circumbinary structure), is balanced. %, and that both the Be disk and circumsecondary region are also stable. 
%Thus, the accretion rate changes according to the amount of material immediately available for accretion, i.e., material that is not in the Be disk, that has not escaped through L3 and L2, and that is not trapped in the circumsecondary region.
Thus, the steady-state accretion rate will be controlled by several factors that set the mass flow rate leaving the Be disk and the fraction of it trapped by the circumstellar structure. All this depends on the model parameters.
Similarly, as seen in Fig.~\ref{fig:densmaps_cont}, the size and density of the circumsecondary region also depends on viscosity and orbital period, as does the amount of matter exiting from the L3 point. For models with $\alpha$ = 0.5 and 1.0, the maximum value of $\dot{M}_{\rm acc}$ is mostly dependent on the period, reaching similar values for both viscosity parameters after the system becomes stable. For models with $\alpha$ = 0.1, this value is systematically lower, and so is the difference between different periods. These are the models that, according to our definition, form an accretion disk. The presence of this rotationally supported structure regulates the accretion rate. }

A counter-intuitive result is that $\dot{M}_{\rm acc}$ decreases with increasing $q$. The same behavior was observed by \citet{franchini2019} in their simulations of misaligned Be binaries. They tentatively attribute this to the fact that a more massive companion could induce stronger torque in the Be disk, thus leading to a stronger truncation and less material entering the RL of the companion. However, our simulations show that while the Be disk is indeed smaller (Fig. \ref{fig:densmaps}), the circumsecondary region is actually larger for higher $q$, and has a strong rotational velocity (rightmost panels of Fig. \ref{fig:grid}, bottom maps in Fig.~\ref{fig:densmaps_cont}). %\fix{A plausible explanation for their lower $\dot{M}_{\rm acc}$ is that, similarly to the trend found for models with $\alpha$ = 0.1, the circumsecondary structure of models 30-1.0-0.33 and 30-1.0-0.50 (being denser and more AM rich than in model 30-1.0-0.16) restricts the accretion rate. }
A plausible explanation for their lower $\dot{M}_{\rm acc}$ is that there is a larger amount of material leaving the system through L2 in these models, as shown in Fig.~\ref{fig:densmaps}.

%\commentACC{Apontar tambem para Fig. 11}
%\ACR{CONTINUE THIS DISCUSSION? IS THE DISK AROUND THE COMPANION MORE VISCOUS, AND THUS EXPANDS -> LEADING TO LESS ACCRETION?}
%\commentACC{Não vejo necessidade. }

%The works of \citet{martin2014} and \citet{franchini2019} show, with SPH simulations using the code \textsc{phantom}, that an accretion disk can indeed be formed when Kozai-Lidov (KL) oscillations are excited in the Be disk by the compact companion in eccentric, non-coplanar orbits. 

% However, an in-depth exploration of accretion in coplanar, non-eccentric Be binaries was not done.

% The layout of our simulations is precisely what makes our results differ from \citet{hayasaki2004} and \citet{martin2014}.     

\subsubsection{Observational expectations}

The accretion of matter from Be decretion disks by their companions is a well-established phenomenon. BeXRBs, comprised of a Be star with a compact companion (a neutron star or black hole), make up half of the high-mass X-ray binary population in the galaxy. Their X-ray emission is consequence of the material of the Be disk falling onto the companion. \revi{There are two types of outbursts in BeXRBs: the weaker, shorter Type I outbursts and the brighter and longer Type II outbursts. Type I outbursts happen when the eccentric orbit binary neutron star accretes material from the Be disk at periastron \citep{okazaki2001, negueruela2001}, while Type II  outbursts are less frequent, not periodic \citep{monageng2017}, and are caused by a large amount of material being suddenly accreted by the companion. This abrupt increase in accretion can be caused by Kozai-Lidov oscillations in the Be disk, excited by a misaligned companion \citep{martin2019}. }

The observational consequences of the accretion will be determined by the nature of the companion. If the companion is a sub-dwarf, there is no direct consequence of the accretion itself on the {observables apart from, perhaps, very faint and soft X-ray emission \citep{naze2022}}. If it is a compact object (a neutron star in particular), strong X-ray emission is expected, as evidenced by the dozens of known BeXRBs. The X-ray luminosity can be written as $L_X = \eta G M_X \dot{M}_{\rm acc}/R_X$, with $\eta=1$ \citep[as per][]{okazaki2001}, and $M_X$ and $R_X$ the mass and radius of the proposed compact object. The values for models with $q = 0.16$ are plotted in Fig. \ref{fig:acrates}, and range from 4 to 7.5 $\times 10^{31}$ \,erg\,$\rm s^{-1}$. For model 30-1.0-0.33, $L_X \approx 1.0 \times 10^{32}$\,erg\,$\rm s^{-1}$, and for model 30-1.0-0.50, $L_X \approx 7.7 \times 10^{31}$\,erg\,$\rm s^{-1}$, respectively the highest and second highest values in all models. All estimates are much lower than the common values of $10^{34}-10^{35}$\,erg\,$\rm s^{-1}$ found for persistent BeXRBs. \fix{Considering the mean accretion rates of $2 \times 10^{-10} \, M_{\odot}/$year found in our models, a companion with a radius of $R_X \approx 0.005$ \req for $q=0.16$, or $R_X \approx 0.01$ \req for $q=0.33$ and $q=0.50$ would reach an X-ray luminosity of $10^{34}$ erg $s^{-1}$. As these radii estimates are about three times larger than the typical size of a WD, and hundreds of times larger than a typical NS, the mean X-ray luminosity of BeXRBs would be easily reached (and surpassed). Thus, in our estimates, a NS or WD in such an orbital configuration would emit strongly in X-rays.} However, our models are coplanar and circular, an unlikely configuration for a BeXRB with a NS since when the NS is formed, the supernova kick could have an impact on the orbit. How significant its impact depends on the previous evolution of the system, the amount of mass ejected during supernova, and the magnitude and direction of the subsequent kick \citep[we point to the ongoing discussion over the formation of BeXRB SGR 0755-2933 in][]{richardson2023, larsen2024, richardson2024}. If the kick does indeed disturb the orbit significantly, the system would not have time to circularize and flatten again in the lifetime of the Be star \citep[see][for the discussion over proposed Be+black hole binary LB-1]{liu2019}.

%\commentACC{Legal este ponto. Tem referencia?} \answerACR{procurando} 

{Recent X-ray observations of Be+sdO binaries show detection of faint X-rays of the same order of magnitude as our models \citep{naze2022}, but the authors note that they do not find a correlation between the X-ray luminosities and the periods of the systems or the properties of the companions. This result does not exclude that this X-ray emission might be coming from accretion onto the companion, or from disk-wind interactions in its vicinity. }
% \commentACC{comparar o vallor de 0.005 que voce estimou com uma WD ou NS}
%\commentACC{Amanda, uma leitura destes valores é que as nossas taxas de acrescao são até maiores do que é necessário para explicar a luminosidade. Ou seja, se a NS deve ser maior, entao nossa taxa de acrecao eh bem maior que o necessário, certo?}
%

The circumsecondary disk might also be directly studied by observations. As discussed on the observational expectations for the bridge (Sect.~\ref{subsec:bridgeobs}), the companion will irradiate this region, and its consequences will depend on the density and kinematics of the material. {Evidence of a material around the secondary was seen in Be binaries $\pi$\,Aqr \citep{bjorkman2002} and, more recently, $o$\,Pup \citep{miroshnichenko2023}.}
A very clear observational example of emission lines from a disk around a companion in a Be binary is HD\,55606, a Be+sdO system. \citet{chojnowski2018} detected strong double peaked \ion{He}{I} lines whose radial velocity shifts are consistent with the companion. The double peaked nature of the emission also indicates the material is rotating, thus suggesting a disk around the sdO. Their Fig. 9 shows their schematic view of the system, similar to \citet{peters2016}'s for HR2142, but without invoking the mass stream crossing L1. The authors assume the Be disk extends past its RL to explain the \ion{O}{I} 8446\,\AA\xspace lines, and attribute the transient shell phases of the Balmer lines, the \ion{O}{I} 7771-7775\,\AA\xspace triplet, and occasionally in the Paschen series, \ion{Si}{II}, \ion{Fe}{II}, and \ion{Ni}{II}, to the denser parts of the spiral dominated disk. 
HD\,55606 is a circular and coplanar binary, with a period of $93.76 \pm 0.02$ days and $q = 0.14 \pm 0.02$, quite similar to our models with \porb = 84 days and $q=0.16$. These models (as all our models, in fact) have the same basic structure, shown in Figs. \ref{fig:mapao} and \ref{fig:densmaps}, and simplified in Fig. \ref{fig:esquema}: the Be disk extends past its RL only in certain regions, there is always a bridge connecting Be disk and companion, and there is always a circumsecondary structure. Future works on the complex radiative transfer taking place in our simulations will answer whether the emission in \ion{O}{I} 8446\,\AA\xspace and the shell features can also originate from the bridge (as suggested by \citealt{peters2016}) or the circumsecondary structure.
% \ACR{Remove this paragraph below? Now this is no longer an issue as the circumsecondary disk is quite symmetric.}
% \commentACC{Ao contrário, podemos escrever que as observacoes agora são apoiadas pelo modelo}
% \editACC{The double peaked, symmetric \ion{He}{I} line profiles of
% HD\,55606 poses a challenge to our models, as they suggest that the circumsecondary is axisymmetric.
% Of our models with similar parameters to HD\,55606 (models 84-x-0.16 - last column of Fig~\ref{fig:densmaps_cont}), none show such symmetry in velocity. 
% However, given the trend in the models of lower viscosity having a more disk-like, rotationally supported structure, it is probable that simulations with $\alpha < 0.1$ will result in more symmetric circumsecondary disks.}
{
Additionally, the double peaked, symmetric \ion{He}{I} line profiles of HD\,55606 are qualitatively supported by our models, that suggest that roughly aximmetric circumsecondary structures are formed (see, e.g., models 84-x-0.16 - last column of Fig.~\ref{fig:densmaps_cont}).
}

Another observable that has provided interesting insight into Be binaries is interferometry. The recent work of \citet{klement2024}, for instance, looks at optical/near-IR interferometry from CHARA\revi{/MIRC(-X)} for 37 Be stars with spectroscopic signs of binarity. HR2142 was one of their targets, for which they also had VLTI/GRAVITY spectrointerferometry for the Br$\gamma$ line. The standard geometric model used in interferometry to describe Be stars and their disks consists of a uniform disk representing the Be star, a 2D Gaussian for continuum disk emission and a Keplerian rotating disk for line emission. For Be binaries, another uniform disk is added to represent the companion. A companion introduces distortions and asymmetries in the disk, which are seen in the interferometric observables, mainly the visibility and the closure phase. The visibility is sensitive to the geometrical size of the source in the sky, while the closure phase is more sensitive to asymmetries in the emission. In order to reproduce the GRAVITY data of HR2142, an unresolved emission line component had to be added to the model, which their best fit placed in the vicinity of the sdO companion, thus confirming the presence of circumsecondary material.

Future studies are planned to investigate the interferometric signals of the models presented in this paper, and how these signals could be used to detect new Be binaries through interferometry or to characterize the circumstellar properties of known ones. \revi{There are currently several interferometric instruments in operation that combined would provide a very complete overview of the structure and dynamics of the Be disk: CHARA/SPICA (covering the visible band, including H$\alpha$), CHARA/MIRC-X (near-infrared bands J and H), VLTI/GRAVITY (near-infrared K band, including Br$\gamma$), and VLTI/MATISSE (mid-infrared L, M and N bands, including Br$\alpha$). CHARA/MIRC-X has already been used to confirm several binaries,
and VLTI/GRAVITY have already proved its cabability in detecting signals from the circumsecondary disk. CHARA/SPICA could conceivably detect the spiral structure in the Be disk in addition to material around secondary, both of which should have big impacts on H$\alpha$.
Finally, in the near future, the upgraded GRAVITY+ will be able to observe fainter Be stars with higher spectral resolution, greatly increasing our sample size of observable Be binaries. }

% By giving the density and velocity outputs of our SPH simulations to the 3D NLTE radiative transfer code \textsc{hdust} \citep{carciofi2006b, carciofi2008b}, we calculated synthetic interferometric images of our simulations in the Br$\gamma$ line, mimicking an observation of ESO VLTI GRAVITY. Using the spectro-interferometric data modeling module \texttt{PMOIRED} \citep{merand2022}, we attempted to fit our synthetic data to the standard Be star model. 

\subsection{Circumbinary disk}\label{sec:circumbi}

Circumbinary disks around multiple systems are important and active agents in star and planet formation, in mass transferring systems, and in post-asymptotic giant branch binary systems \citep{kluska2022}. In the particular case of binary Be stars, the material of this circumbinary disk comes from the the Be disk, not from winds or envelope ejection. The possibility of circumbinary disks in Be binaries was first suggested by \citet{peters2016} for HR2142, based on the gaps that protoplanets form in protoplanetary disks. The SPH simulations of BeXRBs of \citet{franchini2019} show the formation of this type of structure around their misaligned, eccentric systems, which suffer from strong Kozai-Lidov oscillations. \fix{The recent work of \citet{martin2024} explores the dependence of the Be disk size with the disk scale height, and also shows the formation of a circumbinary structures. 
However, our work is the first to fully explore the region beyond the ``truncation'' of the Be disk and its observational consequences with detail.}
%\commentACC{podemos dizer que são os primeiros a estudar estas regioes com resolução suficiente?} \answerACR{para binárias coplanares e circulates, com certeza}

As shown in Fig.~\ref{fig:mapao}, the circumbinary disk is formed from matter that escapes from the companion through the L2 point, and matter that escapes the Be gravitational well through L3 {(as the system is rotating, exit points are not exactly at L2 and L3, but around them)}. While L2 is a exit point for all simulations, L3 can be 
``clogged'' by the resonant torque of the companion when $q$ increases: already at $q = 0.33$ there is virtually no mass loss from L3 in our \porb = 30 days model, as shown in Fig.~\ref{fig:densmaps}. The material in the circumbinary disk has rotational and radial velocities of the order of hundreds of $\rm km\,s^{-1}$, but is bound gravitationaly to the system in the radial extent we considered in our simulations. 
% \commentACC{O que quer dizer com connected? bound?}
Given more time to evolve and grow, with the constant injection of matter and AM coming from the Be star, the circumbinary disk could extend sufficiently to escape the system. 

% \ACR{Plot for this section? Maybe velocity maps similar to Fig. 8? Or contours like Fig. 9? Is it necessary?}
% \commentACC{Não senti falta, mas voce pode apondar o leitor para o mapao e para a Fig.8}

\subsubsection{Observational expectations}

Though much less dense than the Be disk, the bridge, and the circumsecondary structure, the circumbinary disk has non-negligible density and may have observational counterparts. In \citet{klement2017, klement2019}, the authors find that many Be stars have an SED turndown, an effect caused by disk truncation. However, for many Be stars, the shape of the SEDs is inconsistent with a true truncation of the disk by a companion. The slope of the SED after this truncation radius indicates that there is some material beyond the orbit of the companion emitting in IR and radio. The authors find this evidence of a circumbinary disk for all their targets with sufficient data. Therefore, we anticipate that the nature of the circumbinary disks observed in our simulations could be studied using radio photometry or even interferometry.

% \ACR{interferometria? seria possível resolver em rádio?} \commentACC{pus a frase acima, bem sem pretencao.}

%
%______________________________________________________________

\section{Temporal evolution}
\label{sec:temporal_evol}

The fact that Be disks form from decretion rather than accretion is their most distinguishing characteristic. Our SPH simulations build the disk from scratch: they begin with only the sink particles that represent the Be star and the companion, and the gas particles that compose the disk are added continuously as the simulation evolves. As such, each of the regions described in the previous sections appear sequentially, from the inner disk to the circumbinary disk, on a timescale primarily dictated by viscosity. The viscous diffusion timescale is given by
\begin{equation}
    % t_{\rm diff} = \frac{V_{\rm crit}}{\alpha c_s^2} \, r^{1/2} \,. 
    t_{\rm diff} = \frac{r^2}{\alpha c_s H}
    \label{eq:visctimescale}
\end{equation}
\noindent Thus, the timescale for the build up of the disk is lower for larger viscosity (larger $\alpha$), or, simply put, more viscous disks grow faster \citep{haubois2012}.

In our simulations, mass loss is happening constantly, but for most Be stars, the mass ejection mechanism turns on and off. Therefore, the disks of real Be stars are, in the majority of cases, intermittent, being built and dissipated in time scales of months to decades on the whims of the Be mass loss mechanism.  

The volatility of Be disks hinders our ability to infer the presence of a companion via its effects on the disk as discussed throughout this work. A certain period of time is required for the Be disk to grow enough to reach the companion (see Table \ref{tab:inner}), and only then can the bridge, circumsecondary region, and circumbinary disk be formed. Thus, if we observe a system where the Be star has only just started to form a disk, the companion would not have affected it significantly enough yet, and the observational signals discussed in this work would not be present, at least not in their full potential. For instance, in our simulations with 50 day periods and $\alpha = 0.5$ and 1.0, about 500 days of constant mass ejection from the Be star are necessary to produce a fully formed inner Be disk and spiral dominated disk (see Table \ref{tab:inner}), and the other structures would still require even more time to be formed. 

Figure~\ref{fig:buildup} shows the number of orbits necessary for a given region of the system (a volume of $\Delta r$ = 1 \req, $\Delta \phi$ = 0.2 rad, and $\Delta z$ = 4\req, {centered on $\phi, z = 0$}) to reach a quasi steady-state density for models 50-0.5-0.16 and 50-1.0-0.16. For the lower $\alpha$ model (50-0.5-0.16 -- left panel in the figure), the inner disk becomes stable relatively fast ($\approx$5 orbits - 250 days), but the density in the circumsecondary region (for this model, the secondary is at 25.6 \req\,-- pink line in the figure) takes nearly 20 orbits to flatten out. The circumbinary region beyond it (yellow line in the figure) follows closely, also stabilizing after 20 orbits. As per Eq.~\ref{eq:visctimescale}, the more viscous model 50-1.0-0.16 grows and stabilizes faster, with all regions stable after $\approx$10 orbits (500 days).

% \begin{table*}
%     \centering
%     \caption{Estimated stabilization times for key locations in the binary system.}
%     \label{tab:stable}
%     \begin{tabular}{llllll}
%         \hline
%          Model name & Inner Be disk & 5\req before secondary & Position of the secondary & 5\req after secondary & 15\req after secondary \\
%          \hline
%          30-0.1-0.16 & & & & & \\

%          \hline
%     \end{tabular}
% \end{table*}

To ascertain whether a disk is truly fully built up and thus subject to strong influence by a companion, regular monitoring of the Be star is necessary. Spectroscopy is particularly interesting, as the Balmer emission lines are the clearest indicators of disk development, and can be acquired by amateur astronomers with modest equipment for the brightest targets. Amateurs have contributed immensely to Be star databases such as \textsc{BeSS}\footnote{\url{http://basebe.obspm.fr}}. 

\fix{Although more slowly than their build-up, Be disks also dissipate in relatively short timescales \citep{rimulo2018}. When the injection of mass and AM from the central star stops, the disk dissipates inside-out \citep{haubois2012}. Therefore, the inner Be and spiral dominated disk will be the first regions to be affected: as they are depleted, their contribution in the observables also diminish. The circumsecondary region, on the other hand, will take longer to feel the effects of the halting of mass and AM injection. As the matter around the secondary is bound to it, and will eventually be accreted. If the system is observed at this point in its evolution, the detection of the circumsecondary material (and of the secondary itself), is more likely, as its observable contribution is no longer drowned out by the Be disk. In their observations of the Be binary $\pi$ Aqr during a phase of disk dissipation, \citet{bjorkman2002} detected radial velocity shifts in H$\alpha$ that were not compatible with the motion of the Be disk, but rather from a source outside of it. The authors attribute the emission to a cloud of gas around the companion, i.e., the circumsecondary region. Observing dissipating Be disks that had previously been fed for a long period of time offers, therefore, the opportunity to detect faint companions in Be systems. }

\begin{figure}
    \centering
    \includegraphics[width=0.95\linewidth]{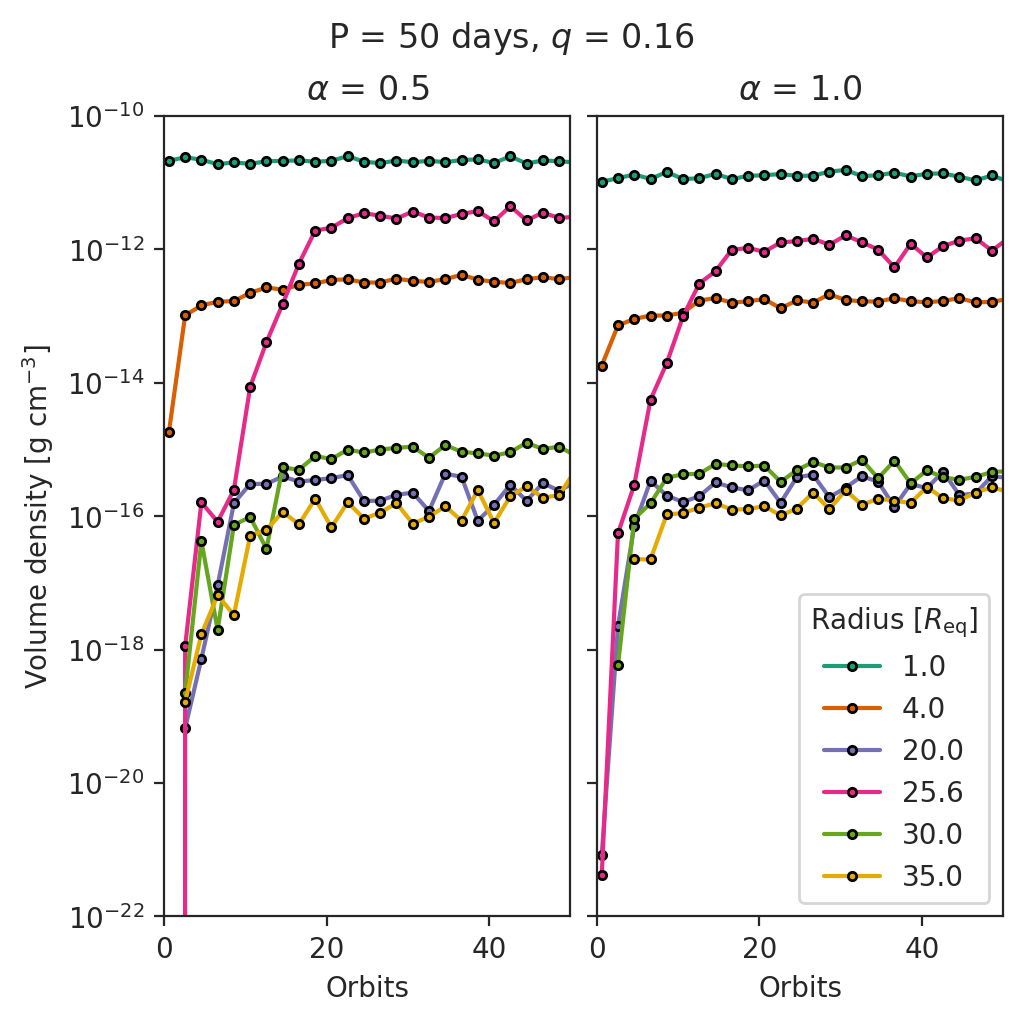}
    \caption{Volume density variation in time in a given volume of $\Delta r$ = 1 \req, $\Delta \phi$ = 0.2 rad, and $\Delta z$ = 4\req, centered on $\phi, z = 0$) for models 50-0.5-0.16 and 50-1.0-0.16.}
    \label{fig:buildup}
\end{figure}

\section{Conclusions}\label{sec:conclusions}

While interacting binary systems are well studied in literature, binary Be stars are unique systems given the nature of their decretion disks. In this work, we detail the structure and kinematics of Be stars in coplanar, circular binary systems using SPH simulations with unprecedented resolution, exploring orbital periods, mass ratios and disk viscosity. 

We identify five regions of the system with different characteristics and observational signatures:
\begin{itemize}
    \item Inner Be disk: innermost region of the decretion disk around the Be star, whose size depends strongly on the period and viscosity parameter of the disk. For less viscous disks, the accumulation effect is very significant; this denser disk would lead to stronger emission in lines that are formed in this region, such as Fe lines, as well as large polarization and infrared excess levels.
    \item Spiral dominated disk: the influence of the secondary on the kinematics of the disk leads to the formation of a double armed density wave, more prominent and tightly wound the less viscous the disk. The observational effects of the waves are V/R variations in emission lines, in particular H$\alpha$ and H$\beta$. The amplitude variation of the density waves decreases with binary separation, which likely leads to a correlation between V/R amplitude with orbital period, such as recently found by \citet{miroshnichenko2023}. 
    %\ACR{falar do paper do Anatoly?}.
    %\commentACC{Sim.}
    \item Bridge: the Be disk is not completely truncated by the tidal interaction with the secondary. What takes place is in fact more similar to Roche lobe overflow, where the decretion disk fills the lobe of the Be star, and matter is transferred to the lobe of the companion. The bridge is this connection between the two lobes, an extension of the leading spiral arm. Observationally, we expect to see asymmetries and features in absorption or emission lines, in particular in shell stars.
    \item Circumsecondary region: matter that enters the RL of the companion is partially accreted and partially lost to the circumbinary region, escaping through the L2 point. The accretion rate is higher the smaller the relative velocity of the material that enters the RL to the companion. Depending on the density of the accretion disk of the companion, emission lines can be excited, which will behave differently from the Be disk lines. All of our simulations form disk-like (i.e., symmetric in density and rotationally supported) structure in this region. 
    %In our simulations, only models with $\alpha = 0.1$ and model 30-1.0-0.50 (which has higher $\alpha$ e larger $q$) form a disk-like structure in this region. 
    \item Circumbinary spiral: outside of the orbit of the companion, material that escapes the system through the L2 and L3 points spirals together in a large one armed wave that encompasses the whole system. Although the spiral has very low density, continuum emission in radio wavelengths (and possibly in the far IR) is possible. 
\end{itemize}

\revi{This five-region structure was predicted observationally by \citet{peters2016} and by the SPH simulations of \citet{panoglou2016}, but this is the first work to show these structures forming in an SPH simulation of a Be binary system from first principles. The circumsecondary region has been confirmed by spectroscopic and/or interferometric observations to exist in a handful of systems (HR\,2142 -- \citet{peters2016, klement2024}, HD\,55606 -- \citet{chojnowski2018}, $o$\,Pup -- \citet{miroshnichenko2023}, $\pi$ Aqr -- \citet{bjorkman2002}, and V658\,Car -- de Amorim, priv. comm.). 
The bridge and circumbinary spiral are less constrained.}
Our simulations confirm that these complex structures can be responsible for peculiar observational behavior of some known binary Be stars. The velocities of the absorption features seen in the intermittent shell phases of Be+sdO HR2142 are compatible with what we see in the bridge in our models. The complex features seen in the spectra of Be+sdO HD55606 might also come from the bridge or from the circumsecondary disk. 
Phase-dependent traveling features like these, observed in both absorption and emission in other Be stars, may indicate binarity according to our results. \revi{Radio excess found in Be binaries that also present SED turndown is a tentative indicator of the presence of diffuse material around the system, likely the circumbinary spiral seen in our simulations.}

The significance of the regions described here for the spectroscopic appearance of the system depends on the evolutionary state of the disk, which is influenced by the viscosity, the variability of the Be star mass ejection, and the duration of the outburst phase. For a given orbital period and viscosity parameter, it can take years for the bridge, circumsecondary and circumbinary structures to be formed. Thus, any attempt to observe the expected observational effects we describe should take the disk evolution into account.

In an expansion of this work, we plan to \revi{include radiative transfer calculations with \textsc{hdust} to our SPH models. This will provide synthetic photometry, spectroscopy, polarimetry and interferometry data that can be directly compared to observations. }
%explore the interferometric and spectroscopic signals of our SPH models, as these observables are important tracers of the various regions described here. 
This planned study will provide constraints for the observational detection and characterization of binary Be stars. %\revi{Our next steps also include implementing our SPH simulations in radiative transfer calculations with \textsc{hdust}, which will provide synthetic photometry, spectroscopy, polarimetry and interferometry data that can be directly compared to observations. }

% - TALK ABOUT HD55606 AND TAJAN'S STAR: THE CIRCUMSECONDARY REGION\\
% - TALK ABOUT THE RADIO OBSERVATIONS OF BINARY BE's: THE CIRCUMBINARY SPIRAL\\

% - INTERFEROMETRY

\begin{acknowledgements}
    A.C.R. acknowledges support from the 'Coordenação de Aperfeiçoamento de Pessoal de Nível Superior' (CAPES grant 88887.464563/2019-00), 'Deutscher Akademischer Austauschdienst' (DAAD grant 57552338), the ESO Studentship program (in particular Dr. Dietrich Baade and Dr. Antoine Mérand), and the Max Planck Institut für Astrophysik in Garching, Germany.  
      A.C.C. acknowledges support from CNPq (grant 314545/2023-9) and `Fundação de Amparo à Pesquisa do Estado de São Paulo' (FAPESP grants 2018/04055-8 and 2019/13354-1). J.E.B. acknowledges support from NSF grant AST-1412135.
      C.E.J. acknowledges support through the National Science and Engineering Research Council of Canada. M.W.S. acknowledges support via the Ontario Graduate Scholarship program
      T.H.A. acknowledges support from FAPESP (grant 2021/01891-2) and CAPES (grant 88887.834998/2023-00).
      This study was granted access to and greatly benefited from the HPC resources of: 
      Centro de processamento de Dados do IAG/USP (CPD-IAG), whose purchase was made possible by the Brazilian agency FAPESP (grants 2019/25950-8, 2009/54006-4).
      IDRIS under the allocation 2023-A0150414439 (resp. A. Domiciano de Souza) made by GENCI; Simulations Intensives en Géophysique, Astronomie, Mécanique et Mathématiques (SIGAMM) infrastructure (cluster Licallo), hosted by Observatoire de la Côte d'Azur (crimson.oca.eu) and supported by the Provence-Alpes Côte d'Azur region, France; National Laboratory for Scientific Computing (LNCC/MCTI, Brazil) of the SDumont supercomputer;

    %\fix{To all: please send your acknowledgements}
\end{acknowledgements}

%-------------------------------------------------------------------
\bibliographystyle{aa}
\bibliography{bib}
% \begin{thebibliography}{}

%   \bibitem[1966]{baker} Baker, N. 1966,
%       in Stellar Evolution,
%       ed.\ R. F. Stein,\& A. G. W. Cameron
%       (Plenum, New York) 333

%    \bibitem[1988]{balluch} Balluch, M. 1988,
%       A\&A, 200, 58

%    \bibitem[1980]{cox} Cox, J. P. 1980,
%       Theory of Stellar Pulsation
%       (Princeton University Press, Princeton) 165

%    \bibitem[1969]{cox69} Cox, A. N.,\& Stewart, J. N. 1969,
%       Academia Nauk, Scientific Information 15, 1

%    \bibitem[1980]{mizuno} Mizuno H. 1980,
%       Prog. Theor. Phys., 64, 544
   
%    \bibitem[1987]{tscharnuter} Tscharnuter W. M. 1987,
%       A\&A, 188, 55
  
%    \bibitem[1992]{terlevich} Terlevich, R. 1992, in ASP Conf. Ser. 31, 
%       Relationships between Active Galactic Nuclei and Starburst Galaxies, 
%       ed. A. V. Filippenko, 13

%    \bibitem[1980a]{yorke80a} Yorke, H. W. 1980a,
%       A\&A, 86, 286

%    \bibitem[1997]{zheng} Zheng, W., Davidsen, A. F., Tytler, D. \& Kriss, G. A.
%       1997, preprint
% \end{thebibliography}

\end{document}